\documentclass[runningheads]{llncs}
\usepackage[T1]{fontenc}
\usepackage{times}  
\usepackage[edges]{forest}
\usepackage[edges]{forest}
\usepackage[edges]{forest}

\forestset{
  roof/.style={
    for children={no edge},
    tikz+={
      \path (.children first) coordinate (firstchild);
      \path (.children last) coordinate (lastchild);
      \draw (.south) -- (firstchild.north);
      \draw (.south) -- (lastchild.north);
      \draw (firstchild.north) -- (lastchild.north);
    }
  }
}

\usepackage{amssymb}
\usepackage{amsmath}
\DeclareUnicodeCharacter{22C6}{\ensuremath{\star}}
\usepackage{microtype}
\usepackage{listings}
\usepackage{helvet}  
\usepackage{courier}  
\usepackage[hyphens]{url}  
\usepackage{graphicx} 
\usepackage[numbers]{natbib}  
\usepackage{caption} 
\usepackage{amsmath}
\usepackage{graphicx}
\usepackage{amssymb}
\usepackage{float}
\usepackage{algorithm}
\usepackage{algorithmic}

\newfloat{dataset}{t}{lod}
\floatname{dataset}{Dataset element}
\usepackage{tabularx}
\usepackage{booktabs}
\newcolumntype{Y}{>{\centering\arraybackslash}X} 
\usepackage{comment}
\usepackage{pgfkeys}
\usepackage{tikz}
\usetikzlibrary{
  positioning,
  shapes.geometric,
  arrows.meta,
  backgrounds,
  fit
}

\usepackage{newfloat}
\usepackage{listings}
\DeclareCaptionStyle{ruled}{labelfont=normalfont,labelsep=colon,strut=off} 
\floatstyle{ruled}
\newfloat{listing}{tb}{lst}{}
\floatname{listing}{Listing}

\usepackage[utf8]{inputenc}
\usepackage{xcolor}
\usepackage{enumitem}
\usepackage{adjustbox}
\usepackage[most]{tcolorbox}
\tcbuselibrary{skins,breakable}
\usepackage[normalem]{ulem}
\usepackage{subcaption}

\definecolor{QP1c}{RGB}{30,80,200}
\definecolor{QP2c}{RGB}{190,40,10}
\definecolor{Trc}{RGB}{110,110,110}
\definecolor{Hdc}{RGB}{20,40,80}
\definecolor{BG1}{RGB}{218,230,255}
\definecolor{BG2}{RGB}{255,226,218}
\definecolor{Ellc}{RGB}{120,40,140}
\definecolor{RDc}{RGB}{20,140,60}
\definecolor{LDc}{RGB}{190,100,0}
\definecolor{RNRc}{RGB}{0,130,150}
\definecolor{BG3}{RGB}{215,245,220}
\definecolor{BG4}{RGB}{255,238,210}
\definecolor{BG5}{RGB}{205,242,245}
\newcommand{\QPA}[1]{{\color{QP1c}\bfseries #1}}
\newcommand{\QPB}[1]{{\color{QP2c}\bfseries #1}}
\newcommand{\Tr}[1]{{\color{Trc}\ensuremath{t_{#1}}}}
\newcommand{\atl}[1]{\ensuremath{\langle\!\langle #1 \rangle\!\rangle}}
\newcommand{\RDA}[1]{{\color{RDc}\bfseries #1}}
\newcommand{\LDA}[1]{{\color{LDc}\bfseries #1}}
\newcommand{\RNRA}[1]{{\color{RNRc}\bfseries #1}}

\forestset{
  xt/.style={
    for tree={
      parent anchor=south,
      child anchor=north,
      align=center,
      font=\footnotesize,
      inner sep=2pt,
      outer sep=0pt,
      l sep=10pt,
      s sep=3pt,
      edge={thin, black}
    }
  },
  ft/.style={
    for tree={
      parent anchor=south,
      child anchor=north,
      align=center,
      font=\small,
      inner sep=2pt,
      outer sep=0pt,
      l sep=12pt,
      s sep=10pt,
      edge={thin, black}
    }
  }
}
\graphicspath{{../../images/}{images/}{./}}

\begin{document}
\title{Translating Natural Language to Strategic
Temporal Specifications via LLMs}
%
%
\author{Marco Aruta\inst{1} \and Francesco Improta \inst{2} \and Vadim Malvone \inst{2} \and Aniello Murano \inst{1} \and Vladana Perlić \inst{2,3}}
\authorrunning{M. Aruta et al.}
%
\institute{University of Naples Federico II, Naples, Italy \and LTCI, Télécom Paris, Institut Polytechnique de Paris, Palaiseau, France \and STIMicroelectronics, Grenoble, France}
\maketitle              
\begin{abstract}
A rigorous formalization of system requirements is a fundamental prerequisite for the verification of Multi-Agent Systems (MAS). However, writing correct formal specifications is well known as an error-prone, time-consuming, and expertise-intensive task. This difficulty is further accentuated in MAS, where requirements must capture strategic abilities and temporal objectives. At present, there is no established methodology for deriving MAS specifications from natural language. We present a framework for translating Natural Language descriptions of strategic requirements into well-formed ATL/ATL$^*$ formulas using Large Language Models (LLMs). Since no available dataset supports supervised learning for the NL-to-ATL/ATL$^*$ translation task, we create and curate a novel expert-validated dataset, employed for training and evaluating fine-tuned models. On a held-out test set, evaluated under the LLM judge that best agrees with expert annotations, in-domain fine-tuning of small open-weight models ($3$ - $7$B parameters) matches strong few-shot proprietary API baselines. Our best fine-tuned system reaches $0.84$ semantic accuracy, statistically on par with $0.86$ for the strongest few-shot proprietary baseline, while keeping requirements on-premises. We further find that judge reliability is inverse to generator strength. The open-weight Llama-3.3-70B tracks human verdicts most closely, whereas the strongest proprietary models are the least reliable judges, over-rejecting faithful paraphrases of the reference. To assess the practical applicability of the generated specifications, we embed our tool to an existing strategic logics model checker, enabling non-expert users to specify strategic properties in natural language.
\end{abstract}

\section{Introduction}

Formal verification of Multi-Agent Systems (MAS) requires logical formalisms capable of capturing how multiple agents interact, compete, and cooperate over time \cite{shoham2008multiagent,wooldridge2009introduction,DBLP:journals/tocl/MogaveroMPV14}. In this setting, strategic logics such as Alternating-time Temporal Logic (ATL$^\star$) \cite{alur2002alternating} provide a natural framework for specifying what coalitions of agents can enforce, regardless of the behavior of others. This makes the logic particularly suitable for safety-critical and adversarial scenarios, where guarantees must hold even in the presence of uncontrollable or competing entities. Yet, writing correct ATL$^\star$ specifications requires substantial expertise in formal methods, temporal logic, and game-theoretic semantics, making the construction of strategic specifications difficult to non-expert practitioners. This remains true even for ATL, the vanilla fragment of ATL$^\star$. A major source of difficulty is that requirements are typically expressed in natural language, which is inherently ambiguous, both because its interpretation depends on context and because the same surface form may admit different structural readings \cite{saussure1916course,small2013lexical}. In formal verification, however, ambiguity cannot be tolerated: verification pipelines require well-formed formulas with precise syntax and unambiguous semantics. More generally, the gap between linguistic expression and logical form has long motivated the view that the surface grammar of a sentence may fail to make its underlying formal structure explicit \cite{frege1892sense,russell1905denoting,Russell1919-RUSTPO-65}. In the context of strategic reasoning, this mismatch becomes especially critical, since ambiguities may affect coalition attribution, temporal scope, and the intended interpretation of alternative strategic outcomes \cite{kulkarni2025dynamic}. These issues are further amplified when natural-language requirements are processed by Large Language Models (LLMs). Besides inheriting the intrinsic ambiguity of natural language, LLM inputs may also be under-specified or contain irrelevant task-related details, both of which can lead to multiple plausible formalizations of the same request. This is particularly problematic in the strategic setting, where even small differences in interpretation may produce logically distinct ATL or ATL$^\star$ formulas with substantially different verification outcomes.

Recent advances in LLMs have generated growing interest in the automatic translation of natural-language requirements into formal specifications. Existing work, however, has focused predominantly on single-agent or purely temporal formalisms, such as Linear Temporal Logic (LTL) \cite{pnueli1977temporal}, Signal Temporal Logic (STL) \cite{maler2004monitoring}, and Computation Tree Logic (CTL) \cite{clarke1981design}. By contrast, the strategic setting of MAS remains largely unexplored. To the best of our knowledge, there is currently no established trained LLM, nor any publicly available dataset, specifically designed to support the translation of natural-language requirements into ATL or ATL$^\star$ specifications.

\paragraph{Our Contributions.}
In this paper, we take a first step toward bridging this gap by introducing an open-weight and open-source LLM framework for translating natural language requirements involving strategic reasoning into well-formed ATL/ATL$^{*}$ formulas. Our contributions are the following:

\begin{itemize}
    \item \textbf{First expert-authored NL-to-ATL/ATL$^\ast$ dataset.}
    We release, to the best of our knowledge, the first benchmark specifically designed for translating natural-language requirements into ATL/ATL$^\ast$ formulae. The dataset contains 1100 instances, hand-written by two domain experts, covering coalitional ability, adversarial outcomes, temporal objectives, and five targeted ambiguity families: right dislocation, left dislocation, VP ellipsis, Right Node Raising, and quantifier scope ambiguity. The dataset is approximately balanced with respect to both the ATL/ATL$^\ast$ temporal operators and the targeted ambiguities types and uses diverse natural-language realizations to avoid surface-template bias. 

    \item \textbf{Hybrid open-weight translation framework.}
    We propose an LLM-based framework for mapping natural-language requirements to well-formed ATL/ATL$^\ast$ specifications. The framework supports both Azure OpenAI inference and private local deployment with open-weight models, enabling use in both large-scale cloud settings and privacy-sensitive industrial environments.

    \item \textbf{Fine-tuning and a judge-reliability finding.}
    We fine-tune open-weight models on the proposed dataset and compare them against strong few-shot proprietary baselines, showing that in-domain fine-tuning of small open-weight models matches few-shot proprietary systems under the judge that best agrees with human experts, while improving transparency and deployment control. That judge is itself open-weight: we find judge reliability to run \emph{opposite} to generator strength, with the open-weight Llama-3.3-70B tracking human verdicts best and the strongest proprietary models judging worst by over-rejecting input-grounded paraphrases, even of their own outputs.

    \item \textbf{Verification-oriented integration.}
    We integrate the framework into \textsc{VITAMIN}~\cite{DBLP:conf/ifaamas/0001M25}, enabling natural-language strategic requirements to be translated into ATL/ATL$^\ast$ formulae and immediately used within an existing model-checking workflow.
    
\end{itemize}
We release the dataset, models, prompts, evaluation scripts, and orchestration code under an open-source license, providing a reusable foundation for research on LLM-assisted formal specification of Multi-Agent Systems.

\paragraph{Related works.} 
The problem of translating natural language specifications into temporal logics has received increasing attention in recent years \cite{cosler2023iterative,Finkbeiner2020TeachingTL,nayak2022req2spec,wu2022autoformalization}, particularly in the context of Linear Temporal Logic (LTL), Signal Temporal Logic (STL), and Computation Tree Logic (CTL). To the best of our knowledge, the most relevant peer-reviewed contributions along this research line are ~\cite{cosler2023nl2spec,fuggitti2023nl2ltl,DBLP:conf/icse/HeBNIG22,zrelli5334216automatic,mendoza2024translating}, which represent the current state of the art in mapping natural-language requirements to temporal logic specifications. The Fuggiti's and Cosler's works \cite{fuggitti2023nl2ltl,cosler2023nl2spec} primarily focus on LTL and explore different degrees of natural language abstraction and compositionality, while He et al. \cite{DBLP:conf/icse/HeBNIG22} and Zrelli et el. \cite{zrelli5334216automatic} respectively focus on STL and CTL. To the best of our knowledge, no prior peer-reviewed work has systematically addressed the problem of translating unrestricted natural language specifications into strategic temporal logics that explicitly capture agent coalitions and their strategic interaction. In this sense, our work constitutes the first dataset-driven and tool-supported study targeting natural language translation into ATL/ATL$^\ast$, thus extending existing approaches beyond purely temporal reasoning toward multi-agent strategic reasoning. Against Fugitti and Chakraborti work \cite{fuggitti2023nl2ltl}, we contend that the proposed natural language specifications remain overly aligned with the syntax and semantics of the target translation logic (LTL in that case). By contrast, our approach adopts natural language specifications with a higher expressive grade, allowing a wider range of semantic synonyms and syntactic realizations (e.g. left and right peripheral displacements, right node raising, verbal phrase ellipsis and quantifier scope ambiguities), adding complexity to the previously cited researches. Furthermore, while we acknowledge that the proposed translation mechanism can be valuable for progressively validating translation correctness in a compositional fashion, it arguably presupposes users who are already familiar with recursive construction over well-formed formulas and the associated composition rules. Therefore, our tool is explicitly designed for users who are illiterate in formal logic, namely, who lack prior training in formal methods notation and reasoning. The synthesis of correct LTL formulas from natural language specifications benefits our work, as the already cited literature in this area provides well established mappings for temporal operators. However, LTL lacks the syntactic and semantic machinery required to capture strategic behaviour, which is expressible in ATL/ATL$^{*}$. Accordingly, a central challenge of this work is to enable robust translation of such strategic operators, and their diverse natural language renderings.

\paragraph{Outline.}Section~2 introduces preliminaries for the reader. Section~3 presents the dataset. Section~4 describes the proposed translation framework. Section~5 reports on evaluation and applications. Section~6 reports on the experiments. Section~7 concludes the work.
\section{Background}

We briefly provide notions about 
natural language ambiguities, Concurrent Game Structures, ATL and ATL$^*$ (analyzing both syntax and semantics). We assume some familiarity with standard concepts from temporal logic and multi-agent systems, focusing on the essential aspects for our framework.

\paragraph{Natural Language Ambiguities.}
Linguistics Formal Syntax and Semantics offer well defined formalizations for identifying natural language ambiguity drawing on graph theory and logical analysis \cite{chomsky1970remarks}. We focus on syntactic ambiguities in translation, as they can be reliably detected and verified. Semantic ambiguities arising from subtle differences in the meaning of terms or expressions present a greater challenge. They appear in the dataset but determining the correct interpretation is non trivial. Consequently, accuracy metrics for syntactic ambiguities are more reliable than those for semantic ones (see the Evaluation section for more details). Here we address the ambiguities covered in our work. The first two types are  right and left dislocation, which occurs when a constituent, wether an argument or adjunct appears outside the clause boundaries, either preceding or following it. Right dislocation is illustrated by "They can eventually ensure the game ends, the players", while "The vending machine, Mary can ensure it will work" exemplifies left dislocation. 
\begin{figure}[h]
\centering
\scriptsize
\begin{forest}
for tree={
    grow'=south,
    parent anchor=south,
    child anchor=north,
    align=center,
    l sep=1pt,   
    s sep=1pt     
}
[S
  [S
    [NP [They]]
    [VP
      [V [can\\eventually\\ensure]]
      [CP [the game\\ends]]
    ]
  ]
  [{NP$_{\mathrm{disl}}$}
    [the players]
  ]
]
\end{forest}
\caption{Right Dislocation. Cartographic analysis provided in Appendix A.}
\end{figure}
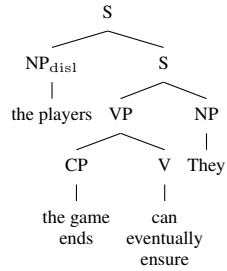

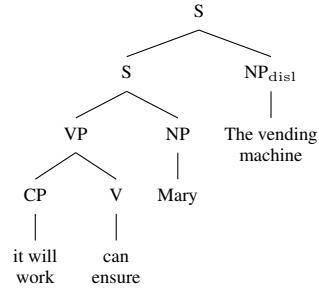
\begin{figure}[h]
\centering
\scriptsize
\begin{forest}
for tree={
    grow'=south,
    parent anchor=south,
    child anchor=north,
    align=center,
    l sep=10pt,
    s sep=8pt
}
[S
  [{NP$_{\mathrm{disl}}$}
    [The vending\\machine]
  ]
  [S
    [NP [Mary]]
    [VP
      [V [can\\ensure]]
      [CP [it will\\work]]
    ]
  ]
]
\end{forest}
\caption{Left Dislocation. Cartographic analysis provided in Appendix A.}
\end{figure}
The third type is Verb Phrase ellipsis (VP ellipsis), a phenomenon whereby a verb phrase is left out of a sentence when its meaning can be recovered from the phrasal context. This typically happens when omitted material mirrors a verb phrase already introduced earlier in the discourse; an example follows "Amanda has a strategy to surely sell all the toys and Leo does too". The fourth type is Right Node Raising (RNR), a construction in which a shared argument appears at the right periphery of a coordinate structure, as in "Bob has, but Albert has not, a strategy to ensure that the vending machine is always operative". The fifth and final type is Quantifier Scope Ambiguity (QSA), where a sentence like "Every agent can prevent a breach" admits two readings. The first is "For every agent, there is some breach that the agent can prevent (possibly a different breach for every agent)". The second is "There is one breach such that every agent can prevent that breach". The respective logical forms in First Order Logic follows:
\begin{itemize}
    \item[-] First reading: $\forall x \, \bigl(A(x) \rightarrow \exists y \, (B(y) \land P(x, y))\bigr)$
    \item[-] Second reading: $\exists y \, \bigl(B(y) \land \forall x \, (A(x) \rightarrow P(x, y))\bigr)$
\end{itemize}
with $A$ being the monadic predicate "is an agent", $B$ the monadic predicate "is a breach", and $P$ the dyadic relation "can prevent", where $P(x,y)$ is read "$x$ can prevent $y$". Since it may not be immediately evident, how scope ambiguities relate to well-formed formulas in logic for strategic reasoning is discussed in the dataset section.

\begin{figure}[h]
\centering
\scriptsize
\begin{forest}
for tree={
    grow'=south,
    parent anchor=south,
    child anchor=north,
    align=center,
    l sep=1pt,
    s sep=1pt
}
[S
  [CoordP
    [S$_1$
      [NP [Bob]]
      [VP [has]]
    ]
    [Coord'
      [Conj [but]]
      [S$_2$
        [NP [Albert]]
        [VP [has not]]
      ]
    ]
  ]
  [{NP$_{\mathrm{shared}}$}
    [a strategy to ensure\\the machine is\\operative]
  ]
]
\end{forest}
\caption{Right Node Raising. Note that traces in VP are omitted for simplicity, same reasoning for the explosion of the complex NP "a strategy to ensure". Cartographic analysis provided in Appendix A.}
\end{figure}
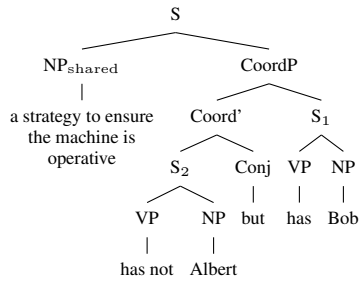
\paragraph{Alternating-time Temporal Logic.}
Alternating-time Temporal Logic (ATL) and ATL$^*$ are strategic temporal logics
for reasoning about the abilities of coalitions of agents in multi-agent systems.
Both logics are interpreted over Concurrent Game Structures (CGS), the standard semantic models for logics for strategic reasoning. Formally, a CGS is a tuple
$G = (Agt, Ap, Act, St, s_0, d, \delta, \ell)$, where:
\begin{itemize}
  \item[-] $Agt$ is a finite, non-empty set of agents;
  \item[-] $Ap$ is a finite, non-empty set of atomic propositions;
  \item[-] $St$ is a non-empty set of states;
  \item[-] $s_0 \in St$ is the initial state;
  \item[-] $Act$ is a non-empty set of actions;
  \item[-] for each agent $a \in Agt$, $d_a : St \to 2^{Act} \setminus \{\emptyset\}$ gives the actions available to agent $a$;
  \item[-] $\delta : St \times \prod_{a \in Agt} Act \to St$ is the transition function;
  \item[-] $\ell : St \to 2^{Ap}$ is the labelling function.
\end{itemize}
Let $\lambda$ be a path, that is, a infinte sequence of states connected by successive transitions. We introduce the following notation:
\begin{itemize}
\renewcommand{\labelitemi}{-}
    \item $\lambda[i]$ denotes the state at position $i$ along $\lambda$, with indexing starting from $0$;
    \item $\lambda[i \dots j]$ denotes the subpath of $\lambda$ from position $i$ to position $j$;
    \item $\lambda[i \dots \infty]$ denotes the suffix of $\lambda$ starting at position $i$.
\end{itemize}
Additionally, we define a history $h \in St^+$ as a finite sequence of states that can occur in the system.
A perfect-recall strategy for agent $a$ is a function
$s_a : St^+ \to Act$ such that $s_a(h) \in d_a(\mathsf{last}(h))$ for every finite history $h$.
A collective strategy for a coalition $A \subseteq Agt$ is denoted by $s_A$.
The set of outcome paths starting from a state $q$ under $s_A$ is denoted by
$\mathsf{out}(q,s_A)$.

\paragraph{Syntax of ATL.}
The following syntax defines a well-formed formula in ATL:
\[
\varphi ::= p \mid \neg \varphi \mid \varphi \wedge \varphi
\mid \langle\!\langle A\rangle\!\rangle X \varphi
\mid \langle\!\langle A\rangle\!\rangle G \varphi
\mid \langle\!\langle A\rangle\!\rangle (\varphi \ U \ \varphi)
\]
Here, $p \in Ap$ and $A \subseteq Agt$. Additional boolean and temporal operators can be derived as usual. Temporal operator $F$ is classically derived.

\paragraph{Semantics of ATL.}
Given a CGS $M$ and a $q \in St$, the satisfaction relation $M,q \models \varphi$ is defined inductively as follows:
\begin{itemize}
  \item[-] $M,q \models p$ iff $p \in \ell(q)$;
  \item[-] $M,q \models \neg \varphi$ iff not $(M,q \models \varphi)$;
  \item[-] $M,q \models \varphi_1 \wedge \varphi_2$ iff
        $M,q \models \varphi_1$ and $M,q \models \varphi_2$;
  \item[-] $M,q \models \langle\!\langle A\rangle\!\rangle X \varphi$ iff
        $\exists s_A \ \forall \lambda \in \mathsf{out}(q,s_A):
        M,\lambda[1] \models \varphi$;
  \item[-] $M,q \models \langle\!\langle A\rangle\!\rangle G \varphi$ iff
        $\exists s_A \ \forall \lambda \in \mathsf{out}(q,s_A) \ \forall i \ge 0:
        M,\lambda[i] \models \varphi$;
  \item[-] $M,q \models \langle\!\langle A\rangle\!\rangle (\varphi_1 \ U \ \varphi_2)$ iff
        $\exists s_A \ \forall \lambda \in \mathsf{out}(q,s_A)$ such that
        $\exists i \ge 0$ with $M,\lambda[i] \models \varphi_2$ and
        $\forall j < i,\ M,\lambda[j] \models \varphi_1$.
\end{itemize}

\paragraph{Syntax of ATL$^*$.}
ATL$^\star$  distinguishes between state and path formulas, defined by:
\[
\begin{array}{l}
\varphi ::= p \mid \neg \varphi \mid \varphi \wedge \varphi
\mid \langle\!\langle A\rangle\!\rangle \gamma, \\
\gamma ::= \varphi \mid \neg \gamma \mid \gamma \wedge \gamma
\mid X\gamma \mid \gamma \ U \ \gamma .
\end{array}
\]
Here, $p \in AP$ and $A \subseteq Agt$. Temporal operators $G$ and $F$ are classically derived.

\paragraph{Semantics of ATL$^*$.}
Semantics of ATL$^*$ extends that for ATL as follows: 
\begin{itemize}
  \item[-] $M,q \models \langle\!\langle A\rangle\!\rangle \gamma$ iff
        $\exists s_A \ \forall \lambda \in \mathsf{out}(q,s_A):
        M,\lambda \models \gamma$;
  \item[-] $M,\lambda \models \varphi$ iff $M,\lambda[0] \models \varphi$;
  \item[-] $M,\lambda \models X\gamma$ iff
        $M,\lambda[1 \dots \infty] \models \gamma$;
  \item[-] $M,\lambda \models \gamma_1 \ U \ \gamma_2$ iff
        $\exists i \ge 0$ s.t.
        $M,\lambda[i \dots \infty] \models \gamma_2$ and
        $\forall j < i,\ M,\lambda[j \dots \infty] \models \gamma_1$.
\end{itemize}

\section{Dataset}
\label{sec:dataset}
The dataset contains a balanced distribution of 1100 elements comprehensing simple sentences, namely those most syntactically aligned with well-formed formulas in ATL/ATL$^\ast$, and sentences exhibiting the four targeted ambiguity types, which we briefly summarize here. Right/Left Dislocation occurs when a constituent (argument or adjunct) appears outside its clause boundaries, either preceding or following it. Verb Phrase (VP) Ellipsis is a phenomenon in which a verb phrase is omitted when its meaning is recoverable from context, typically because the elided material mirrors a verb phrase introduced earlier in the discourse. Right Node Raising (RNR) is a coordination construction in which a shared argument appears at the right periphery. Quantifier Scope Ambiguity (QSA) arises when a sentence contains multiple quantifiers whose relative scope is not uniquely determined by the surface syntax, yielding two or more distinct readings that differ in the logical relationships between the quantified expressions. The dataset constitutes a gold standard resource, that is, a manually annotated benchmark validated by human experts and intended as a reference for evaluating automated systems. It has been reviewed and validated by three annotators with expertise in formal verification and formal logic for strategic reasoning in multi-agent systems. The complete annotation process and inter-annotator discussion for each dataset entry are documented in the supplementary materials. Reviewers highlighted potential issues and abstained from commenting on elements they deemed unproblematic; such abstentions were interpreted as implicit acceptance rather than missing judgments. Validation decisions were aggregated via majority vote, and relative rationales were explicitly recorded. 

\paragraph{Dataset construction, balancing, and ground truth.}
Our dataset was hand-written by two experts in formal verification and the temporal logics for strategic reasoning in multi-agent systems, who manually authored each natural-language specification together with its corresponding ATL or ATL$^\ast$ formalization. All formulae in the dataset are well-formed ATL$^\ast$~ formulae. Since ATL is a syntactic fragment of ATL$^\ast$~, formulae that happen to be expressible in ATL are included as special cases without separate classification. The instances cover a broad range of multi-agent system scenarios, including settings in which goals are achieved by a single strategic agent, by explicitly defined coalitions, or by coordinated behaviour among multiple agents. The dataset was designed to be approximately balanced with respect to the temporal operators appearing in the original definition of ATL and ATL$^\ast$~\cite{alur2002alternating}. In particular, the final distribution of operator occurrences is: $X$: 590, $G$: 590, $F$: 598, $U$: 586. Moreover, we deliberately varied the natural-language realizations of each temporal operator in order to reduce surface-level template bias. For instance, the next operator $X$ was expressed through formulations such as ``at the next step'' and ``in the immediately following state''; eventuality $F$ through expressions such as ``sooner or later'' and ``in due course''; until $U$ through expressions such as ``up to the moment when'' and ``before''; and globally $G$ through expressions such as ``at all times'' and ``in every state''. In addition to temporal balancing, the dataset explicitly targets the aforementioned five linguistic ambiguities. These ambiguity types were introduced according to an approximately uniform distribution within the ambiguous subset, with each type accounting for about 20\% of the targeted ambiguous examples. Right and left dislocation were used to test whether the model can correctly recover the syntactic and semantic role of displaced constituents. VP ellipsis was used to test whether omitted predicates can be reconstructed from the preceding clause. RNR was introduced to test whether the model can recover a shared right-peripheral argument in a coordinate structure, as in cases where two agents differ in their strategic ability but share the same formal objective. QSA was used to test whether the model can distinguish distributive readings from collective coalition-based readings. For this latter class, whenever two interpretations are genuinely admissible, the ground truth includes two formal outputs, corresponding to the distributive and collective readings. The dataset is provided in \textit{JSON} format, following common practice in recent NLP and AI dataset releases, where JSON or JSONL is used to serialize structured instances and associated metadata in a portable and easily processable form~\cite{wang2025verifiable}. Each dataset element is represented as a JSON object containing the natural-language \textit{input} and an \textit{outputs} field. The latter contains the correct\textit{formula} translation of the natural language sentence, encoding the intended strategic and temporal meaning. In particular, for QSA, the same field contains multiple formula pairs, one for each admissible interpretation. All formulae were manually curated and checked to serve as the ground truth for evaluating natural-language-to-ATL/ATL$^\ast$ translation. After describing the structure of the dataset, we report two illustrative ambiguous instances. Dataset element~\ref{ds:vp_ellipsis_example} shows a case of VP ellipsis in a robotic system specification. Here, the phrase ``the rover can too'' does not explicitly repeat the predicate introduced in the first clause. A correct translation therefore requires reconstructing the omitted strategic-temporal content and assigning it to the second agent as well. Dataset element~\ref{ds:qsa_example} shows a case of quantifier scope ambiguity (QSA), the most challenging family: the universally quantified subject ``every rover'' licenses two distinct admissible readings: a \emph{distributive} one, in which each agent individually enforces the objective, and a \emph{collective} one, in which the agents enforce it jointly as a coalition. Both readings are stored as gold, so a correct translation must emit both, since neither alone captures the intended meaning.

\begin{dataset}[t]
\caption{VP ellipsis ambiguity example}
\label{ds:vp_ellipsis_example}
\begin{algorithmic}[1]
\STATE \textbf{ID:} \texttt{ex331}
\STATE \textbf{Input (Natural Language):}
\STATE \quad \textit{``The drone can guarantee that it will return to base sooner or later, and the rover can too.''}

\STATE \textbf{Outputs:}
\STATE \quad
\begin{minipage}[t]{0.92\linewidth}
\textbf{Formula:}
\[
\langle\!\langle \text{Drone} \rangle\!\rangle
F\,\text{at\_base}
\ \wedge\
\langle\!\langle \text{Rover} \rangle\!\rangle
F\,\text{at\_base}
\]
\end{minipage}
\end{algorithmic}
\end{dataset}

\begin{dataset}[t]
\caption{Quantifier scope ambiguity example}
\label{ds:qsa_example}
\begin{algorithmic}[1]
\STATE \textbf{ID:} \texttt{ex336}
\STATE \textbf{Input (Natural Language):}
\STATE \quad \textit{``Every rover can guarantee that it will never enter a hazardous area.''}

\STATE \textbf{Outputs} (both readings are jointly required):
\STATE \quad
\begin{minipage}[t]{0.92\linewidth}
\textbf{Distributive reading} (each agent individually):
\[
\begin{array}{l}
\langle\!\langle \text{Rover}_1 \rangle\!\rangle G\,\neg\text{in\_hazardous\_area\_1} \ \wedge\\
\langle\!\langle \text{Rover}_2 \rangle\!\rangle G\,\neg\text{in\_hazardous\_area\_2} \ \wedge\\
\langle\!\langle \text{Rover}_3 \rangle\!\rangle G\,\neg\text{in\_hazardous\_area\_3}
\end{array}
\]
\end{minipage}

\STATE \quad
\begin{minipage}[t]{0.92\linewidth}
\textbf{Collective reading} (the agents as a coalition):
\[
\langle\!\langle \text{Rover}_1,\text{Rover}_2,\text{Rover}_3 \rangle\!\rangle G\,\neg\text{in\_hazardous\_area}
\]
\end{minipage}
\end{algorithmic}
\end{dataset}

\section{Framework}
\label{sec:framework}

We present an open-source framework that translates natural-language strategic requirements into well-formed ATL/ATL$^\ast$ specifications. It comprises three components: a \emph{hybrid inference} layer that runs local open-weight models and cloud APIs behind a single interface; \emph{scalable, reproducible} experimentation with multi-seed aggregation and cluster execution; and \emph{post-prediction evaluation} that couples syntactic and meaning-level assessment and feeds a strategic model checker. Figure~\ref{fig:tool_pipeline} sketches the architecture; the end-to-end translation-and-evaluation loop is given as Algorithm~1 in the technical appendix.

\subsection{Architecture}
Requests enter through a command-line interface or a REST endpoint and reach an \textbf{Experiment Orchestrator}. It draws data from a \textbf{Dataset Manager} (which produces deterministic, stratified train/validation/test partitions and holds the curated few-shot exemplars out of every split, so no prompting example leaks into evaluation) and resolves models through a \textbf{Model Abstraction Layer}, a declarative registry mapping a short model key to its provider, checkpoint, decoding parameters, and fine-tuning targets. The backing \textbf{Inference Engine} serves local open-weight checkpoints and cloud endpoints under one calling convention; because fine-tuning depends on initialization, each fine-tuned configuration is repeated over several seeds, while greedy decoding keeps run-to-run variation attributable to training rather than sampling. Every generation is stored as a \emph{raw prediction} with its metadata (latency, token counts, decoding configuration) before any scoring, and two evaluators then consume it: an \textbf{Exact-Match Evaluator} for normalized syntactic comparison and an \textbf{LLM-as-a-Judge} for meaning when surface forms diverge. Decoupling inference from evaluation, and applying the same two-tier scoring to local and cloud predictions, ensures that observed differences reflect model capability rather than the harness.

\usetikzlibrary{arrows.meta, positioning, fit, backgrounds, calc}
\begin{figure}[H]
\centering
\resizebox{0.82\columnwidth}{!}{%
\begin{tikzpicture}[
    font=\small\sffamily,
    >=Stealth,
    box/.style={
        rectangle, draw=black!75, thick, rounded corners=3pt,
        align=center, fill=white, minimum height=9mm, inner sep=5pt
    },
    io/.style={box, fill=gray!8},
    core/.style={box, fill=blue!8},
    data/.style={box, fill=yellow!12},
    infer/.style={box, fill=teal!6},
    eval/.style={box, fill=orange!10},
    grp/.style={
        rectangle, draw=black!40, thick, dashed, rounded corners=6pt,
        inner xsep=10pt, inner ysep=8pt
    },
    arr/.style={->, thick, draw=black!80},
    trainarr/.style={->, thick, dashed, draw=teal!90!black},
    line/.style={thick, draw=black!80}
]


\node[core, text width=2.4cm] (orch) at (0, 3) {\textbf{Experiment}\\\textbf{Orchestrator}};
\node[io, text width=2.1cm] (dm) at (-3.1, 3) {\textbf{Dataset}\\Manager};
\node[io, text width=2.1cm] (mal) at (3.1, 3) {\textbf{Model}\\Abstraction};

\node[io, text width=2.1cm] (cli) at (-1.5, 4.6) {CLI / REST\\Endpoint};
\node[io, text width=2.2cm] (vit) at (1.5, 4.6) {\textsc{VITAMIN}\\{\scriptsize Model Checker}};

\node[infer, text width=2.4cm] (local) at (-1.7, 1.2) {\textbf{Open-weight} 3--7B\\{\scriptsize \textit{LoRA fine-tuned}}};
\node[infer, text width=2.4cm] (cloud) at (1.7, 1.2) {\textbf{Proprietary API}\\{\scriptsize \textit{frozen (GPT)}}};

\node[data, text width=3.8cm] (pred) at (0, -0.6) {\textbf{Raw Predictions}\\{\scriptsize multi-seed $+$ metadata}};

\node[eval, text width=2.4cm] (em) at (-1.7, -2.4) {\textbf{Exact-Match}};
\node[eval, text width=2.4cm] (judge) at (1.7, -2.4) {\textbf{LLM-as-a-Judge}};

\node[core, text width=4.8cm] (metrics) at (0, -4.2) {\textbf{Aggregated Metrics}\\{\scriptsize accuracy $\cdot$ agreement $\cdot$ efficiency}};

\begin{scope}[on background layer]
    \coordinate (inf_top) at (0, 2.1);
    \node[grp, fill=teal!3, fit=(local)(cloud)(inf_top)] (inf_box) {};
    \node[below=2pt of inf_box.north, font=\bfseries\scriptsize\color{teal!80!black}] {INFERENCE ENGINE};

    \coordinate (eval_top) at (0, -1.5);
    \node[grp, fill=orange!5, fit=(em)(judge)(eval_top)] (eval_box) {};
    \node[below=2pt of eval_box.north, font=\bfseries\scriptsize\color{orange!80!black}] {EVALUATION HARNESS};
\end{scope}


\draw[arr] (cli.south) |- (0, 3.9) -| ([xshift=-0.5cm]orch.north);
\draw[<->, thick, draw=black!80] (vit.south) |- (0, 3.9) -| ([xshift=0.5cm]orch.north);

\draw[arr] (dm.east) -- (orch.west);
\draw[arr] (mal.west) -- (orch.east);

\draw[trainarr] (dm.west) -- ++(-0.35, 0) |- (local.west) 
    node[pos=0.74, above, font=\scriptsize\color{teal!90!black}, inner sep=3pt] {Train split};

\draw[line] (orch.south) -- (0, 2.45);
\draw[arr] (0, 2.45) -| (local.north);
\draw[arr] (0, 2.45) -| (cloud.north);

\draw[line] (local.south) |- (0, 0.3);
\draw[line] (cloud.south) |- (0, 0.3);
\draw[arr] (0, 0.3) -- (pred.north);

\draw[line] (pred.south) -- (0, -1.25);
\draw[arr] (0, -1.25) -| (em.north);
\draw[arr] (0, -1.25) -| (judge.north);

\draw[line] (em.south) |- (0, -3.4);
\draw[line] (judge.south) |- (0, -3.4);
\draw[arr] (0, -3.4) -- (metrics.north);

\end{tikzpicture}%
}
\caption{Tool architecture. A natural-language requirement (from the CLI, a REST endpoint, or the integrated \textsc{VITAMIN} model checker) reaches the \emph{Experiment Orchestrator}, which draws stratified splits and exemplars from the \emph{Dataset Manager} and resolves models through the \emph{Model Abstraction Layer}. Inference is hybrid: open-weight $3$-$7$B models are LoRA fine-tuned and served locally (dashed arrow), while proprietary API models are used frozen, behind one interface. Each generation is stored as a multi-seed raw prediction and scored by two independent evaluators (exact match and an LLM-as-a-judge) before aggregation into accuracy, agreement, and efficiency metrics.}
\label{fig:tool_pipeline}
\end{figure}
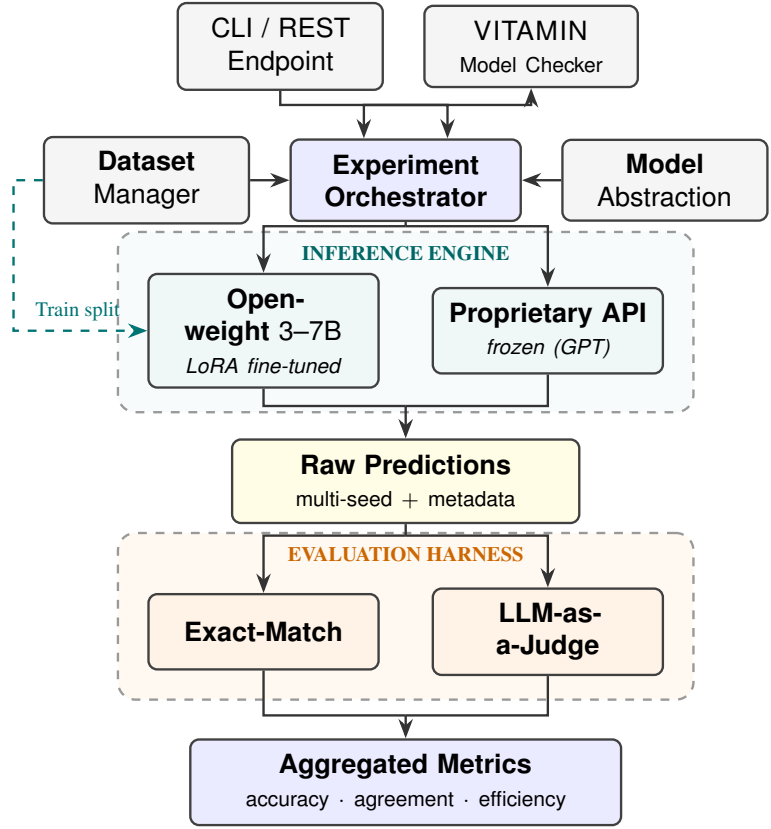

\paragraph{Prompting and Output Parsing.}
\label{subsec:prompting}
The translation target is pinned by a system prompt that fixes the ATL/ATL$^\ast$ surface syntax (the coalition modality $\langle\!\langle A \rangle\!\rangle$, the temporal operators $X$, $F$, $G$, and $U$, the Boolean connectives, \textsc{PascalCase} agent names, and \texttt{snake\_case} atomic propositions) together with the scope convention that keeps the strategic operator separate from the temporal operator it governs, e.g.\ $\langle\!\langle \mathrm{Machine}\rangle\!\rangle\,G(\mathit{paid} \\ \rightarrow \mathit{printed})$ rather than $\langle\!\langle \mathrm{Machine}\rangle\!\rangle(G\,\mathit{paid}\rightarrow \mathit{printed})$. It also states the intended treatment of each targeted ambiguity: repeat the recovered formula for VP ellipsis, share the right-peripheral objective across both conjuncts for Right Node Raising, and emit \emph{all} admissible readings (one per line) for quantifier-scope ambiguity, never fusing two distinct readings into one. The few-shot condition prepends a small curated pool of exemplars spanning these phenomena, held out of every split. Models emit plain text under greedy decoding; a parser then strips extraneous tokens and canonicalizes whitespace, letter case, and operator spelling, so that scoring compares formulas rather than incidental surface form. A requirement with several readings is stored as a set of gold formulas, and a prediction must reproduce the entire set to count as correct. The full system prompt and the curated few-shot exemplars are reproduced in the technical appendix.

\paragraph{Hybrid Inference: Privacy and Capacity.}
\label{subsec:hybrid}
Because requirement confidentiality is central to industrial formal methods, the framework treats local open-weight inference as a primary path: checkpoints are fine-tuned with low-rank adapters~\cite{hu2022lora}, using $4$-bit QLoRA quantization~\cite{dettmers2023qlora} for the larger $7$B models, so domain specialization happens entirely on-premises and no requirement leaves the air-gapped environment. The same interface targets Azure OpenAI endpoints when data sensitivity permits, letting practitioners trade data sovereignty against reasoning capacity without altering the pipeline.

\paragraph{Scalability and Reproducibility.}
\label{subsec:scale}
The framework targets large parameter sweeps over models, prompting conditions, and seeds: each $(\text{model}, \text{condition}, \text{split})$ cell is an independent unit of work dispatched to a SLURM-managed GPU cluster. To separate genuine capability from initialization noise, the orchestrator \emph{decouples} the data-split seed from the training seed and supports two protocols: a single \emph{canonical} stratified split, identical across models, used for the headline numbers reported here, and an optional stratified $k$-fold cross-validation mode, with shared folds, for estimating split-induced variance. Aggregation runs on the appropriate axis (over seeds or folds), yielding means with dispersion rather than single-point estimates.

\subsection{Model-Checker Integration}
\label{subsec:integration}
A translation is useful only if it fits inside a verification workflow. We therefore embed our framework in \textsc{VITAMIN}, an established model checker for strategic logics~\cite{DBLP:conf/ifaamas/0001M25}, so designers express constraints in natural language and obtain formulas that are immediately parsed and verified (a screenshot is in the technical appendix). The integration is dual-service: the verification backend manages the model-checking environment, while translation runs as a FastAPI microservice whose \texttt{/generate} endpoint takes the requirement, a model key, and prompting flags (few-shot, maximum new tokens, an optional fine-tuned adapter). A non-destructive patch installs a thin client that routes ATL requests to the translation service first and falls back to the checker's native generation whenever the service is unavailable or returns an ill-formed result. Because the emitted formulas are parsed by the same front-end used for verification, only well-formed specifications enter model checking, giving immediate syntactic validation of model output.
\section{Evaluation Protocol}
\label{sec:evaluation}
We evaluate translations with a layered protocol that moves from strict syntax, to meaning, to the reliability of the evaluation itself, and finally to deployment efficiency. Every metric below is reported as a mean over independent runs rather than a single measurement. 

LLM decoding and fine-tuning are sensitive to initialization, so each fine-tuned configuration is run over several training seeds on the canonical split, and accuracy is reported as the seed mean with a $95\%$ confidence interval, so the headline numbers reflect capability rather than a single favorable run.

\subsection{Exact Match and Meaning}
For the exact matching, predictions are first compared to the reference under a normalization that removes purely surface differences: whitespace and letter case are normalized, logical operators are mapped to a canonical ASCII form (e.g.\ $\land,\lor,\neg,\to$), and the spelling of coalition names inside $\langle\!\langle\cdot\rangle\!\rangle$ is normalized while propositions are left intact, so that a non-semantic agent renaming is not penalized but a genuine lexical change still counts as a miss. Because a single requirement may admit several jointly required readings (for instance under quantifier-scope ambiguity), exact match demands that \emph{all} gold formulas be produced, not merely one of them. This deterministic check is the syntactic-precision baseline.

\paragraph{LLM-as-a-judge.} A prediction that is \emph{not} an exact match is forwarded to an LLM-as-a-judge, which decides strategic and temporal equivalence to the reference and returns a structured verdict (a \texttt{correct} label and a free-text rationale). To avoid depending on any single model, we benchmark \emph{six} independent LLM judges (DeepSeek-V3.2, GPT-4.1, GPT-5.2, GPT-5.4, and the open-weight Gemma-2-27B and Llama-3.3-70B) under one fixed prompt (version~v1.4), whose adjudication rubric and calibration examples are reproduced in the technical appendix, and validate each against the expert audit below. Headline accuracy is reported under the single most human-aligned judge (the open-weight Llama-3.3-70B), which is moreover \emph{not} among the evaluated systems and so never grades its own outputs. Judge decisions are cached per judge identity, so no judge is queried twice on an identical triple, and the judges remain independent. Reported accuracy is therefore the exact-match rate plus the judge-recovered fraction, a decomposition we state explicitly.

\paragraph{Why an LLM judge rather than formal equivalence.} One might instead decide semantic equivalence of two ATL/ATL$^\ast$ formulas over concurrent game structures, but this is ill-posed here: the atomic propositions and agent names are grounded in the natural-language request rather than in a fixed game structure, so there is no shared model over which to evaluate satisfaction, and a faithful prediction may use a differently named but corresponding proposition or coalition; equivalence for ATL$^\ast$ is also computationally heavy. The judge therefore targets the appropriate notion: whether the prediction preserves the \emph{strategic-temporal intent} of the reference (coalition attribution, the strategic modality and its polarity, the temporal operators, and the scope relations among them) rather than verbatim form, requiring that all admissible readings appear for requirements with multiple readings. We treat these verdicts as estimates and validate them against human judgment next.

\subsection{Reliability of the Evaluation}
To quantify how trustworthy the automatic verdicts are, we compute pairwise Cohen's $\kappa$ and Fleiss' $\kappa$ across judges. Agreement is measured \emph{only} over items that actually reach the judges: deterministic verdicts, exact matches, and empty or unmatched predictions scored as incorrect without an LLM call, are excluded, since including them would trivially inflate agreement.

\paragraph{Human validation.} An expert audit provides ground truth for the judges themselves. From a stratified sample that deliberately over-represents items on which the judges disagree, and excluding predictions the cloud safety filter blocked at generation (which carry no judge output to validate), $599$ predictions are labeled independently by two expert annotators (inter-annotator agreement $99.5\%$, Cohen's $\kappa=0.99$). The annotators then deliberated over their three initial disagreements and reached agreement on two of them; the third resisted consensus and is retained as a genuine unresolved disagreement, excluded from the human-as-reference comparison and leaving $598$ consensus labels. On both reconciled cases the annotators concluded that the prediction \emph{was} faithful, overturning the stricter initial reading that the automated judges had echoed, so here the human audit corrects an over-strict tendency shared by the LLM judges rather than merely confirming them. The single remaining case sits on the boundary of strategic--temporal equivalence, a limit that no labeling protocol, human or automated, can fully avoid. The consensus labels both measure how closely each judge tracks human judgment and produce a robustness ranking in which human labels \emph{override} the judges wherever an item was audited. Because the judges track human judgment unevenly, we report headline accuracy under the most human-aligned judge alone, with a conservative cross-check that averages all six judges.

\subsection{Efficiency, Significance and Transparency}
We characterize deployment efficiency through the accuracy-latency trade-off rather than a single scalar: for every system we report mean per-example latency alongside accuracy and identify the Pareto frontier, the systems for which no alternative is simultaneously faster and more accurate. We avoid monetary cost models, since token prices and amortized GPU-hour rates are provider and hardware specific, whereas accuracy and latency are measured directly. One caveat applies to latency: cloud models are timed end-to-end over the network, so their latency is comparable within the local group but only indicative across the API/local boundary, while accuracy remains comparable throughout.

To establish whether the leading system is genuinely ahead of its closest competitors, we compare systems on shared items with a paired bootstrap confidence interval and a randomization test, rather than the marginal accuracy gap alone. We also disclose every provider content-filter event: the few inputs rejected by the cloud safety filter (at generation or judging time) count as incorrect in the reported accuracy, are listed explicitly, and the ranking is recomputed with them excluded to confirm that the conclusions do not depend on them.

\section{Experiments}
\label{sec:experiments}

\begin{table}[t]
\centering
\small
\setlength{\tabcolsep}{4pt}
\begin{tabular}{@{}llcccc@{}}
\toprule
& & \multicolumn{2}{c}{\textbf{Baseline}} & \multicolumn{2}{c}{\textbf{Fine-tuned}} \\
\cmidrule(lr){3-4}\cmidrule(lr){5-6}
\textbf{Model} & \textbf{Size} & ZS & FS & ZS & FS \\
\midrule
\multicolumn{6}{@{}l}{\textit{Proprietary (Azure API)}}\\
gpt-4.1        & API  & 0.611 & 0.843 & --    & --    \\
gpt-5.4        & API  & \textbf{0.694} & \textbf{0.856} & --    & --    \\
\midrule
\multicolumn{6}{@{}l}{\textit{Open-weight (local)}}\\
mistral-7b     & 7B   & 0.239 & 0.349 & 0.762 & 0.766 \\
qwen-3b        & 3B   & 0.151 & 0.367 & 0.778 & 0.801 \\
phi3           & 3.8B & 0.275 & 0.417 & 0.800 & 0.818 \\
qwen-coder-7b  & 7B   & 0.298 & 0.463 & \textbf{0.839} & \textbf{0.844} \\
\bottomrule
\end{tabular}
\caption{Semantic accuracy under the most human-aligned judge, namely the open-weight \textsc{Llama-3.3-70B}, which tracks the expert audit best among six candidates (Cohen's $\kappa=0.59$; cf.\ Table~\ref{tab:judge_reliability}) shows the seed mean. ZS/FS denote zero-/few-shot prompting; fine-tuning is available only for the open-weight models; best per column in bold; fine-tuned values are means over three training seeds. Under this judge the strongest few-shot proprietary baseline (\textsc{gpt-5.4}, $0.856$) and the top fine-tuned open-weight system (\textsc{qwen-coder-7b}, $0.844$) are statistically tied within the $n{=}218$ sampling noise, so small fine-tuned open-weight models \emph{match} much larger few-shot proprietary ones. A conservative six-judge cross-check appears in the technical appendix.}
\label{tab:accuracy}
\end{table}

We report an empirical study on the held-out test split using the protocol of Section~\ref{sec:evaluation}. This study addresses two implicit research questions: whether LLMs can translate natural language into strategic temporal logic reliably enough to feed a verifier, and how locally deployable open-weight models compare with proprietary baselines.

\paragraph{Setting.}
We evaluate six models across four conditions, yielding $20$ model-condition systems (fine-tuning applies only to the four open-weight models). Two proprietary Azure OpenAI models, \textsc{gpt-4.1} and \textsc{gpt-5.4}, are run zero- and few-shot as strong baselines: being closed-weight, they cannot be adapted within our on-premises setting, and the modest in-domain data makes API-side fine-tuning at their scale impractical, so they serve as strong \emph{few-shot} reference points. Our central comparison thus contrasts in-domain fine-tuning of small open-weight models with few-shot prompting of much larger proprietary ones, rather than model class alone. Four open-weight models (\textsc{mistral-7b}, \textsc{qwen-3b}, \textsc{phi3} (Phi-3.5-mini), and \textsc{qwen-coder-7b}) are run as baselines and after LoRA fine-tuning, in both prompting conditions. All systems share identical prompt templates, few-shot exemplars, and decoding (one greedy decode per query, with no chain-of-thought, self-consistency, or parser-feedback/self-repair scaffolding for any model), with local fine-tuning and inference on NVIDIA A100 GPUs, so the comparison isolates in-domain fine-tuning versus few-shot prompting rather than inference-time engineering applied unevenly. The parser canonicalizes surface form identically before scoring, and the tool's parser-validated fallback (Section~\ref{subsec:integration}) is a model-agnostic deployment feature held out of this controlled comparison; the proprietary baselines thus receive the same best-effort prompting as the open-weight systems, and the reported accuracies are a conservative floor that such scaffolding could lift for all alike. Of the $1100$ instances, the $7$ curated few-shot exemplars are held out of every split; the remaining $1093$ are partitioned by a stratified $70/10/20$ split into $765$ training, $110$ validation, and $218$ test examples. Each fine-tuned configuration is repeated over three training seeds, and every prediction is scored by six LLM judges (DeepSeek-V3.2, GPT-4.1, GPT-5.2, GPT-5.4, and the open-weight Gemma-2-27B and Llama-3.3-70B) plus an expert human audit on a stratified subsample, for $4{,}279$ items with multiple judge verdicts.

\paragraph{Fine-tuning versus few-shot baselines.}
Table~\ref{tab:accuracy} reports semantic accuracy by model and condition under the most human-aligned judge. The top is a statistically tied cluster: the strongest few-shot proprietary baseline, \textsc{gpt-5.4} few-shot ($0.856$), and the best fine-tuned open-weight system, \textsc{qwen-coder-7b} few-shot ($0.844$), differ by only $0.012$, within the $n{=}218$ sampling noise, so small fine-tuned open-weight models \emph{match} much larger few-shot proprietary ones. Fine-tuning is the dominant factor for the open-weight models: it lifts \textsc{phi3} from $0.417$ to $0.818$ in the few-shot setting and moves every open-weight model from weak zero-shot baselines (as low as $0.151$) into the $0.76$--$0.84$ band. Few-shot prompting helps throughout, but its benefit is largest for the untuned baselines and shrinks after fine-tuning, so task adaptation, rather than scale alone, drives open-weight performance: the $3$--$7$B fine-tuned models reach the level of much larger \emph{few-shot} proprietary systems. These gains come from fine-tuning on the $765$-instance training split, augmented twofold by templated paraphrasing of the input (synonym substitutions for temporal and strategic phrasings that leave the gold formula unchanged), applied to training only, so validation and test are never augmented; modest in-domain supervision thus suffices to specialize a small model to ATL/ATL$^\ast$. The $95\%$ seed confidence intervals for the top systems are narrow (about $\pm0.02$; see the six-judge robustness check in the technical appendix), but they capture only training-seed variation at the fixed canonical split and \emph{not} test-set sampling variance, which at $n{=}218$ items is on the order of $\pm0.06$; the seed intervals are therefore a lower bound, and a stratified $k$-fold sweep, which our framework supports, would estimate the split component directly and is left to future runs.

\paragraph{Exact match versus meaning.}
We decompose accuracy under the most human-aligned judge into a deterministic exact-match floor and the fraction the LLM judge recovers (the appendix gives the full breakdown for the headline systems under the same judge). The gap between syntactic and meaning-level scoring is large and uneven, and widest for the proprietary baselines. For \textsc{gpt-5.4} zero-shot, exact match yields only $0.213$ of the $0.694$ accuracy: roughly $69\%$ of correct answers are syntactically distinct from the reference yet strategically equivalent; for \textsc{gpt-4.1} few-shot the judge-recovered share is $46\%$. It is smaller for the strongest fine-tuned systems ($23$ - $29\%$), whose outputs more often match the reference surface form, but never negligible. Purely syntactic metrics therefore systematically understate translation quality, especially on underspecified requirements.

\paragraph{Statistical significance.}
A paired analysis on shared items confirms a tight top cluster and shows that the comparison with the proprietary baselines is itself judge-dependent, in the direction our judge-reliability analysis predicts. The fine-tuned \textsc{phi3} and \textsc{qwen-coder-7b} systems (few- and zero-shot) are mutually indistinguishable (pairwise differences $\le 0.024$ under the six-judge mean, within the $n{=}218$ sampling noise) and far above the untuned open-weight baselines. Against \textsc{gpt-5.4} few-shot the paired difference changes sign with the judge: under the most human-aligned judge of Table~\ref{tab:accuracy} the two are statistically level (\textsc{qwen-coder-7b} few-shot $\Delta=-0.009$, randomization $p=0.71$), whereas under the six-judge mean the open-weight systems edge ahead by a small margin (\textsc{qwen-coder-7b} $\Delta=0.052$, paired bootstrap $95\%$ CI $[0.004,0.102]$, $p=0.03$; \textsc{phi3} $\Delta=0.048$, $p=0.03$). That swing is precisely the artifact documented above (the stricter judges penalise the proprietary baselines' faithful input-grounded paraphrases more heavily, while the most human-aligned judge credits them), so we read the generation comparison as \emph{parity}: the fine-tuned open-weight systems are statistically level with the few-shot proprietary baselines under the most reliable judge, and the small lead that appears under stricter scoring reflects judge over-strictness rather than a true quality gap.

\paragraph{Judge reliability.}
{\sloppy\setlength{\emergencystretch}{3em}
Table~\ref{tab:judge_reliability} summarizes inter-rater reliability, and it carries our most striking result: \emph{judge quality runs opposite to generator strength}. The six LLM judges agree only moderately with one another (mean pairwise Cohen's $\kappa=0.49$) and track the human audit very unevenly. The strongest, most recent proprietary models are the \emph{worst} judges. More specifically, \textsc{Gpt-5.4}, the top few-shot \emph{generator} in Table~\ref{tab:accuracy}, manages only $\kappa=0.24$ against humans and \textsc{gpt-5.2} worse still ($\kappa=0.16$). The best judge by a wide margin is the open-weight \textsc{Llama-3.3-70B} ($\kappa=0.59$, $81.8\%$ agreement), ahead of \textsc{DeepSeek-V3.2} ($0.46$) and \textsc{gpt-4.1} ($0.44$). The cause is systematic over-strictness on predicate \emph{aliasing}: on $1167$ items both strict proprietary judges (\textsc{gpt-5.4} and \textsc{gpt-5.2}) rejected a prediction that \textsc{Llama-3.3-70B} accepted and on the audited subset the human experts sided with \textsc{Llama} over them $160$ to $26$, while there is \emph{not a single} item where \textsc{Llama} rejected a prediction both of these judges accepted. Almost all such cases merely rename an atomic proposition to an input-grounded paraphrase (\texttt{discharged}\,$\to$\,\texttt{patient\_discharged}, \texttt{error}\,$\to$\,\texttt{error\_state}) while preserving every coalition, temporal operator and connective, which the rubric instructs judges to accept but \textsc{gpt-5.4}/\textsc{gpt-5.2} reject as ``not a clearly grounded alias''. This strictness is not self-serving: \textsc{gpt-5.4} as a judge marks down the \emph{proprietary} generations too, scoring its own few-shot output at $0.634$ where \textsc{Llama} gives $0.856$, so the strong proprietary models penalise even their own family and only the open-weight judge recovers their true accuracy. We therefore report headline accuracy under \textsc{Llama-3.3-70B}, the most human-aligned judge and itself external to the evaluated systems. The choice is robust: under the conservative six-judge mean the fine-tuned open-weight systems even \emph{lead} the proprietary baselines, and when human labels override the judges wherever an item was audited the top cluster is unchanged; only the \emph{size} of the open-weight margin moves with the judge, shrinking to a tie under the human-aligned \textsc{Llama}.\par}

\begin{table}[t]
\centering
\small
\begin{tabular}{@{}lccc@{}}
\toprule
\textbf{Judge vs.\ human} & $n$ & \textbf{Agr.} & \textbf{Cohen's} $\kappa$ \\
\midrule
\textbf{Llama-3.3-70B} & $598$ & $\mathbf{81.8\%}$ & $\mathbf{0.59}$~(mod.) \\
DeepSeek-V3.2 & $598$ & $73.7\%$ & $0.46$~(mod.) \\
GPT-4.1       & $598$ & $72.7\%$ & $0.44$~(mod.) \\
Gemma-2-27B   & $598$ & $59.4\%$ & $0.25$~(fair) \\
GPT-5.4       & $598$ & $56.9\%$ & $0.24$~(fair) \\
GPT-5.2       & $598$ & $50.7\%$ & $0.16$~(slight) \\
\bottomrule
\end{tabular}
\caption{\textbf{Judge reliability against the human audit.} Agreement of each of the six LLM judges with the consensus expert labels on the audited items; ``Agr.''\ is the exact label-match rate and the $\kappa$ interpretation follows Landis-Koch. Among the judges themselves, mean pairwise Cohen's $\kappa=0.49$ (range $0.19$-$0.62$; $65$-$91\%$ raw agreement); adding the human as a seventh rater gives Fleiss' $\kappa=0.42$. The open-weight \textsc{Llama-3.3-70B} tracks humans most closely (well ahead of the substantially more recent and more capable \textsc{gpt-5.4} and \textsc{gpt-5.2}) so it serves as the headline judge, illustrating that judge quality runs opposite to generator quality. Each judge is compared against the $598$ audited predictions that carry a consensus expert label; one further audited prediction, on which the annotators reached no consensus, is excluded (its case appears in the technical appendix).}
\label{tab:judge_reliability}
\end{table}

\paragraph{Efficiency.}
We characterize efficiency through the accuracy--latency trade-off, plotted in the technical appendix. The two fastest fine-tuned open-weight systems are also among the most accurate: few-shot \textsc{qwen-3b} ($1567$\,ms) and \textsc{phi3} ($1681$\,ms), both faster than the proprietary baselines \textsc{gpt-4.1} ($1979$\,ms) and \textsc{gpt-5.4} ($2538$\,ms); few-shot \textsc{qwen-coder-7b} reaches the top open-weight accuracy ($3338$\,ms), and fine-tuned \textsc{mistral-7b} is slowest ($4460$\,ms). The accuracy-latency frontier is therefore shared between the fast fine-tuned open-weight systems and the marginally more accurate few-shot proprietary baselines; since the open-weight models match them on accuracy (Table~\ref{tab:accuracy}) while running on-premises and, for the fastest two, at lower latency, they remain preferable for local deployment. The proprietary latencies are network-timed and only indicative across the API/local boundary.
The Azure safety filter blocked only two inputs (a Shakespeare quotation and one other), counted as incorrect; excluding them leaves every system's rank unchanged, so no conclusion hinges on provider-side filtering.

\paragraph{Error analysis by ambiguity type.}
The failure modes cluster by linguistic phenomenon. The dominant errors of the \emph{untuned} open-weight models are syntactic: predictions borrow CTL-style path quantifiers (e.g.\ \texttt{AF\,p}) for the coalition modality, introduce first-order quantifiers and variables foreign to ATL/ATL$^\ast$, or emit malformed coalitions, all rejected by both exact match and the judges. Fine-tuning largely removes these gross errors, accounting for the steep zero-shot-to-fine-tuned jump. The residual errors concentrate on the ambiguity families: under quantifier-scope ambiguity the models most often \emph{collapse} the distributive and collective readings into one formula or return only one of the two required; under VP ellipsis and Right Node Raising they fail to \emph{reconstruct} the shared objective for the second agent; and dislocation occasionally yields a mis-attributed coalition. Temporal-operator substitutions (e.g.\ \texttt{U} for \texttt{F}) and spurious or dropped negations occur across categories but are rare once a model is fine-tuned. These patterns track the benchmark's design intent: once surface syntax is mastered, the discriminating difficulty is recovering the intended \emph{scope} and \emph{distribution} of strategic ability, not lexical translation. The technical appendix makes this quantitative on the $31$ multi-reading (QSA) and $187$ single-reading test items (a re-aggregation of existing verdicts, no new labels). Single-reading accuracy is uniformly high ($0.85$-$0.88$), but every system drops sharply on the multi-reading slice, which remains the hardest phenomenon for all of them. Here no model class dominates: the strongest few-shot proprietary baseline (\textsc{gpt-5.4}, $0.68$) and the best fine-tuned open-weight model (\textsc{qwen-coder-7b}, $0.63$) are comparable, while the smaller fine-tuned models lag (\textsc{qwen-3b} $0.33$, \textsc{mistral-7b} $0.27$); with only $31$ QSA items these differences are within noise (binomial $95\%$ CI $\approx\pm0.17$), so we read the slice as a qualitative contrast rather than a ranking. Under the strictest judges the proprietary baselines instead collapse on this slice, an artifact of their over-strict rejection of input-grounded predicate paraphrases (documented in our judge-reliability analysis) rather than a failure to emit both readings, and one the human-aligned headline judge does not share.

\paragraph{Discussion.}
Taken together, our experiments show that in-domain fine-tuning of small open-weight models translates strategic requirements as accurately as strong few-shot proprietary baselines while keeping requirements on-premises; that open-weight models also make the most reliable automatic \emph{judges} of these translations, whereas the strongest proprietary models judge worst by over-rejecting faithful paraphrases; and that exact match alone is an inadequate metric. Complete tables, the training and decoding configuration, the computing infrastructure, and a reproducibility checklist appear in the technical appendix and supplementary material.

\section{Conclusions}
\label{sec:conclusions}

We introduced an open-weight, open-source framework for translating natural-language strategic requirements into well-formed ATL/ATL$^\ast$ specifications. We provided an expert-curated dataset explicitly designed for the NL-to-ATL/ATL$^\ast$ task and covering both simple requirements and five ambiguity families: right dislocation, left dislocation, VP ellipsis, Right Node Raising, and Quantifier Scope Ambiguity. The dataset is approximately balanced with respect to ATL/ATL$^\ast$ temporal operators and provides a reusable benchmark for training and evaluating models for strategic specification synthesis. We also presented a hybrid framework running both cloud APIs and local open-weight models with LoRA fine-tuning; and a multi-faceted evaluation protocol coupling normalized exact match with a human-validated LLM-as-a-judge. Our experiments show that in-domain fine-tuning of small, locally deployable open-weight models ($3$-$7$B) translates strategic requirements as accurately as strong few-shot proprietary baselines under the most human-aligned judge; that exact match alone understates quality, since faithful translations often differ syntactically from the reference; and that the automatic judges must themselves be validated, since judge reliability runs opposite to generator capability, the open-weight Llama-3.3-70B is the most human-aligned judge, while the strongest, most recent proprietary models are the least reliable, over-rejecting faithful paraphrases even of their own generations. Finally, the integration with \textsc{VITAMIN} shows how natural-language strategic requirements can be connected to existing model-checking workflows.

\paragraph{Limitations.}
Several limitations qualify these results. The benchmark, although expert-validated, is curated rather than drawn from deployed industrial requirements, and is English-only; transfer to in-the-wild specifications and other languages remains to be demonstrated. We target ATL/ATL$^\ast$, so requirements involving knowledge, beliefs, or the more expressive constructs of Strategy Logic lie outside the present scope. Semantic accuracy is measured by an LLM judge whose agreement with human experts (moderate even for the best-aligned judge) is imperfect, so the absolute numbers are best read as calibrated estimates rather than exact rates; we mitigate this with multiple independent judges and a human-overridden ranking. Our reported confidence intervals, moreover, reflect training-seed variation at a single canonical split; with only $218$ test items, test-set resampling adds uncertainty of order $\pm0.06$ that these intervals do not capture, so they are best read as a lower bound, a stratified $k$-fold evaluation, supported by our framework, would tighten this estimate. Finally, equivalence is assessed by strategic-temporal intent rather than by formal model checking, and the reported cross-environment latencies are only indicative across the API/local boundary.

\paragraph{Future Work.}
This seminal study provides a springboard for several future research directions. The main avenue concerns extending the dataset by releasing it publicly to the formal methods community for their contribution, as well as broadening the framework to more expressive strategic formalisms, such as Strategy Logic and its fragments~\cite{DBLP:journals/tocl/MogaveroMPV14,DBLP:conf/aaai/CermakLM15,DBLP:conf/ijcai/BelardinelliJKM19}, or epistemic variants of ATL~\cite{DBLP:journals/fuin/JamrogaH04,aminof2016prompt,jamroga2017reasoning}. This would enable translations involving knowledge, beliefs, and information flow in multi-agent systems. Moreover, the released dataset opens the door to training domain-specialized models potentially reducing reliance on post-hoc semantic validation.


%
%
%
\bibliographystyle{splncs04}
\bibliography{biblio}

@article{kulkarni2025dynamic,
  title={Dynamic coalition structure detection in natural language-based interactions},
  author={Kulkarni, Abhishek N and Liu, Andy and Gaglione, Jean-Raphael and Fried, Daniel and Topcu, Ufuk},
  journal={arXiv preprint arXiv:2502.16339},
  year={2025}
}

@inproceedings{DBLP:conf/ijcai/BelardinelliJKM19,
  author       = {Francesco Belardinelli and
                  Wojciech Jamroga and
                  Damian Kurpiewski and
                  Vadim Malvone and
                  Aniello Murano},
  title        = {Strategy Logic with Simple Goals: Tractable Reasoning about Strategies},
  booktitle    = {Proc. of {IJCAI} 2019},
  pages        = {88--94},
  publisher    = {ijcai.org},
  year         = {2019}
}

@inproceedings{jamroga2017reasoning,
  title={Reasoning about natural strategic ability},
  author={Jamroga, Wojciech and Malvone, Vadim and Murano, Aniello},
  booktitle={Proceedings of the 16th Conference on Autonomous Agents and MultiAgent Systems},
  pages={714--722},
  year={2017}
}

@article{aminof2016prompt,
  title={Prompt Alternating-Time Epistemic Logics.},
  author={Aminof, Benjamin and Murano, Aniello and Rubin, Sasha and Zuleger, Florian},
  journal={KR},
  volume={16},
  pages={258--267},
  year={2016}
}

@inproceedings{DBLP:conf/aaai/CermakLM15,
  author       = {Petr Cerm{\'{a}}k and
                  Alessio Lomuscio and
                  Aniello Murano},
  title        = {Verifying and Synthesising Multi-Agent Systems against One-Goal Strategy
                  Logic Specifications},
  booktitle    = {Proc. of {AAAI}},
  pages        = {2038--2044},
  publisher    = {{AAAI} Press},
  year         = {2015}
}

@article{wang2025verifiable,
  title={Verifiable format control for large language model generations},
  author={Wang, Zhaoyang and Jiang, Jinqi and Zhou, Huichi and Zheng, Wenhao and Zhang, Xuchao and Bansal, Chetan and Yao, Huaxiu},
  journal={arXiv preprint arXiv:2502.04498},
  year={2025}
}

@inproceedings{fuggitti2023nl2ltl,
  title={Nl2ltl--a python package for converting natural language (nl) instructions to linear temporal logic (LTL) formulas},
  author={Fuggitti, Francesco and Chakraborti, Tathagata},
  booktitle={AAAI 2023},
  volume={37},
  number={13},
  pages={16428--16430},
  year={2023}
}

@inproceedings{cosler2023nl2spec,
  title={nl2spec: Interactively translating unstructured natural language to temporal logics with large language models},
  author={Cosler, Matthias and Hahn, Christopher and Mendoza, Daniel and Schmitt, Frederik and Trippel, Caroline},
  booktitle={CAV 2023},
  pages={383--396},
  year={2023},
  organization={Springer}
}

@inproceedings{pnueli1977temporal,
  title={The temporal logic of programs},
  author={Pnueli, Amir},
  booktitle={18th annual symposium on foundations of computer science (sfcs 1977)},
  pages={46--57},
  year={1977},
  organization={ieee}
}

@book{shoham2008multiagent,
  title={Multiagent systems: Algorithmic, game-theoretic, and logical foundations},
  author={Shoham, Yoav and Leyton-Brown, Kevin},
  year={2008},
  publisher={Cambridge University Press}
}

@book{wooldridge2009introduction,
  title={An introduction to multiagent systems},
  author={Wooldridge, Michael},
  year={2009},
  publisher={John wiley \& sons}
}

@article{alur2002alternating,
  title={Alternating-time temporal logic},
  author={Alur, Rajeev and Henzinger, Thomas A and Kupferman, Orna},
  journal={Journal of the ACM (JACM)},
  volume={49},
  number={5},
  pages={672--713},
  year={2002},
  publisher={ACM New York, NY, USA}
}

@inproceedings{clarke1981design,
  title={Design and synthesis of synchronization skeletons using branching time temporal logic},
  author={Clarke, Edmund M and Emerson, E Allen},
  booktitle={Workshop on logic of programs},
  pages={52--71},
  year={1981},
  organization={Springer}
}

@inproceedings{DBLP:conf/ifaamas/0001M25,
  author       = {Angelo Ferrando and
                  Vadim Malvone},
  title        = {{VITAMIN:} VerIficaTion of {A} MultI ageNt system},
  booktitle    = {{AAMAS} 2025},
  pages        = {3023--3025},
  publisher    = {IFAAMAS},
  year         = {2025}
}

@article{zrelli5334216automatic,
  title={Automatic Translation of Natural Language Requirements into Ctl Specifications Using Large Language Models: A Multi-Approach Evaluation⋆},
  author={Zrelli, Rim and Misson, Henrique Amaral and Ben Attia, Marwa and Gohring de Magalhaes, Felipe and Shabah, Abdo and Nicolescu, Gabriela},
  journal={Available at SSRN 5334216}
}

@article{cosler2023iterative,
  title={Iterative circuit repair against formal specifications},
  author={Cosler, Matthias and Schmitt, Frederik and Hahn, Christopher and Finkbeiner, Bernd},
  journal={arXiv preprint arXiv:2303.01158},
  year={2023}
}

@article{Finkbeiner2020TeachingTL,
  title={Teaching Temporal Logics to Neural Networks},
  author={Bernd Finkbeiner and Christopher Hahn and Markus Norman Rabe and Frederik Schmitt},
  journal={ArXiv},
  year={2020},
  volume={abs/2003.04218},
  url={https://api.semanticscholar.org/CorpusID:212633677}
}

@inproceedings{nayak2022req2spec,
  title={Req2Spec: Transforming software requirements into formal specifications using natural language processing},
  author={Nayak, Anmol and Timmapathini, Hari Prasad and Murali, Vidhya and Ponnalagu, Karthikeyan and Venkoparao, Vijendran Gopalan and Post, Amalinda},
  booktitle={{REFSQ} 2022},
  pages={87--95},
  year={2022},
  organization={Springer}
}

@inproceedings{DBLP:conf/icse/HeBNIG22,
  author       = {Jie He and
                  Ezio Bartocci and
                  Dejan Nickovic and
                  Haris Isakovic and
                  Radu Grosu},
  title        = {DeepSTL - From English Requirements to Signal Temporal Logic},
  booktitle    = {{ICSE} 2022},
  pages        = {610--622},
  publisher    = {{ACM}},
  year         = {2022}
}

@article{wu2022autoformalization,
  title={Autoformalization with large language models},
  author={Wu, Yuhuai and Jiang, Albert Qiaochu and Li, Wenda and Rabe, Markus and Staats, Charles and Jamnik, Mateja and Szegedy, Christian},
  journal={Advances in neural information processing systems},
  volume={35},
  pages={32353--32368},
  year={2022}
}

@book{saussure1916course,
  address = {London},
  author = {de Saussure, Ferdinand},
  note = {(trans. Roy Harris)},
  publisher = {Duckworth},
  title = {Course in General Linguistics},
  year = {[1916] 1983}
}

@book{small2013lexical,
  title={Lexical Ambiguity Resolution: Perspective from Psycholinguistics, Neuropsychology and Artificial Intelligence},
  editor={Small, Steven L. and Cottrell, Garrison W. and Tanenhaus, Michael K.},
  publisher={Elsevier},
  year={2013},
  isbn={9780080510132},
  pages={518}
}

@inproceedings{maler2004monitoring,
  title={Monitoring temporal properties of continuous signals},
  author={Maler, Oded and Nickovic, Dejan},
  booktitle={International symposium on formal techniques in real-time and fault-tolerant systems},
  pages={152--166},
  year={2004},
  organization={Springer}
}

@article{russell1905denoting,
  author  = {Russell, Bertrand},
  title   = {On Denoting},
  journal = {Mind},
  volume  = {14},
  number  = {56},
  pages   = {479--493},
  year    = {1905}
}

@article{Russell1919-RUSTPO-65,
	author = {Bertrand Russell},
	doi = {10.5840/monist191929120},
	journal = {The Monist},
	number = {1},
	pages = {32--63},
	title = {The Philosophy of Logical Atomism},
	volume = {29},
	year = {1919}
}

@article{frege1892sense,
  title={On sense and reference},
  author={Frege, Gottlob},
  journal={Translations from the philosophical writings of Gottlob Frege},
  volume={2},
  pages={56--85},
  year={1892},
  publisher={Oxford}
}

@incollection{chomsky1970remarks,
  title={Remarks on Nominalization},
  author={Chomsky, Noam},
  editor={Jacobs, Roderick A. and Rosenbaum, Peter S.},
  booktitle={Readings in English Transformational Grammar},
  pages={184--221},
  year={1970},
  publisher={Ginn},
  address={Waltham, MA}
}

@inproceedings{mendoza2024translating,
  title={Translating natural language to temporal logics with large language models and model checkers},
  author={Mendoza, Daniel and Hahn, Christopher and Trippel, Caroline},
  booktitle={2024 Formal Methods in Computer-Aided Design (FMCAD)},
  pages={1--11},
  year={2024},
  organization={IEEE}
}

@article{DBLP:journals/tocl/MogaveroMPV14,
  author       = {Fabio Mogavero and
                  Aniello Murano and
                  Giuseppe Perelli and
                  Moshe Y. Vardi},
  title        = {Reasoning About Strategies: On the Model-Checking Problem},
  journal      = {{ACM} Trans. Comput. Log.},
  volume       = {15},
  number       = {4},
  pages        = {34:1--34:47},
  year         = {2014},
  url          = {https://doi.org/10.1145/2631917},
  doi          = {10.1145/2631917},
  bibsource    = {dblp computer science bibliography, https://dblp.org}
}

@article{DBLP:journals/fuin/JamrogaH04,
  author       = {Wojciech Jamroga and
                  Wiebe van der Hoek},
  title        = {Agents that Know How to Play},
  journal      = {Fundam. Informaticae},
  volume       = {63},
  number       = {2-3},
  pages        = {185--219},
  year         = {2004},
  url          = {http://content.iospress.com/articles/fundamenta-informaticae/fi63-2-3-05},
  bibsource    = {dblp computer science bibliography, https://dblp.org}
}

@inproceedings{hu2022lora,
  title={{LoRA}: Low-Rank Adaptation of Large Language Models},
  author={Hu, Edward J. and Shen, Yelong and Wallis, Phillip and Allen-Zhu, Zeyuan and Li, Yuanzhi and Wang, Shean and Wang, Lu and Chen, Weizhu},
  booktitle={International Conference on Learning Representations (ICLR)},
  year={2022}
}

@inproceedings{dettmers2023qlora,
  title={{QLoRA}: Efficient Finetuning of Quantized {LLMs}},
  author={Dettmers, Tim and Pagnoni, Artidoro and Holtzman, Ari and Zettlemoyer, Luke},
  booktitle={Advances in Neural Information Processing Systems (NeurIPS)},
  year={2023}
}

@article{fodor1982referential,
  author = {Fodor, Janet Dean and Sag, Ivan A.},
  title = {Referential and Quantificational Indefinites},
  journal = {Linguistics and Philosophy},
  volume = {5},
  number = {3},
  pages = {355--398},
  year = {1982}
}

@article{reinhart1997quantifier,
  author = {Reinhart, Tanya},
  title = {Quantifier Scope: How Labor Is Divided between QR and Choice Functions},
  journal = {Linguistics and Philosophy},
  volume = {20},
  number = {4},
  pages = {335--397},
  year = {1997}
}

@incollection{rizzi1997fine,
  author = {Rizzi, Luigi},
  title = {The Fine Structure of the Left Periphery},
  booktitle = {Elements of Grammar},
  editor = {Haegeman, Liliane},
  pages = {281--337},
  publisher = {Kluwer},
  address = {Dordrecht},
  year = {1997}
}
\appendix
\section{Technical Appendix}

The syntactic trees in this appendix employ X-bar theoretic structures to map natural language sentences to ATL/ATL$^\ast$ formulas. Some trees use representational conventions that prioritize transparent scope and argument structure for translation purposes, rather than strictly derivational syntactic analyses. Therefore, we explicitly flag them as interpretive devices for the syntax-semantics interface. These conventions do not affect the empirical coverage of the test suite, which is designed to probe whether models correctly recover logical form from surface form.

\subsection{Example 1: Quantifier Scope Ambiguity Surface Structure}

\begin{example}[Surface structure]
Consider the sentence:
\begin{quote}
Every rover can guarantee that it will never enter a hazardous area.
\end{quote}
The sentence contains two quantificational expressions: the universal DP \emph{every rover} and the indefinite DP \emph{a hazardous area}. This gives rise to two possible semantic readings, (i) a surface-scope reading and (ii) an inverse-scope reading.
\end{example}

\begin{figure}[t]
\centering
\begin{tcolorbox}[
  enhanced,
  arc=4pt,
  boxrule=1pt,
  colback=yellow!12,
  colframe=orange!70!black,
  title={\bfseries Quantifier Scope Ambiguity --- Surface Structure},
  fonttitle=\normalsize
]
\centering
\begin{adjustbox}{max width=\linewidth}
\begin{forest} xt
[CP
  [C$'$
    [C [$\varnothing$]]
    [TP
      [\QPA{DP$_i$}
        [D$'$ [D [every]] [NP [N [rover]]]]
      ]
      [T$'$
        [T [can]]
        [VP
          [V$'$
            [V [guarantee]]
            [CP
              [C$'$
                [C [that]]
                [TP
                  [DP [it]]
                  [T$'$
                    [T [will]]
                    [NegP
                      [Neg$'$
                        [Neg [never]]
                        [VP
                          [V$'$
                            [V [enter]]
                            [\QPB{DP$_j$}
                              [D$'$ [D [a]] [NP [AP [hazardous]] [N [area]]]]
                            ]
                          ]
                        ]
                      ]
                    ]
                  ]
                ]
              ]
            ]
          ]
        ]
      ]
    ]
  ]
]
\end{forest}
\end{adjustbox}
\end{tcolorbox}
\caption{Surface structure for a quantifier scope ambiguity involving a universal DP and an indefinite DP.}
\label{fig:qsa-surface}
\end{figure}
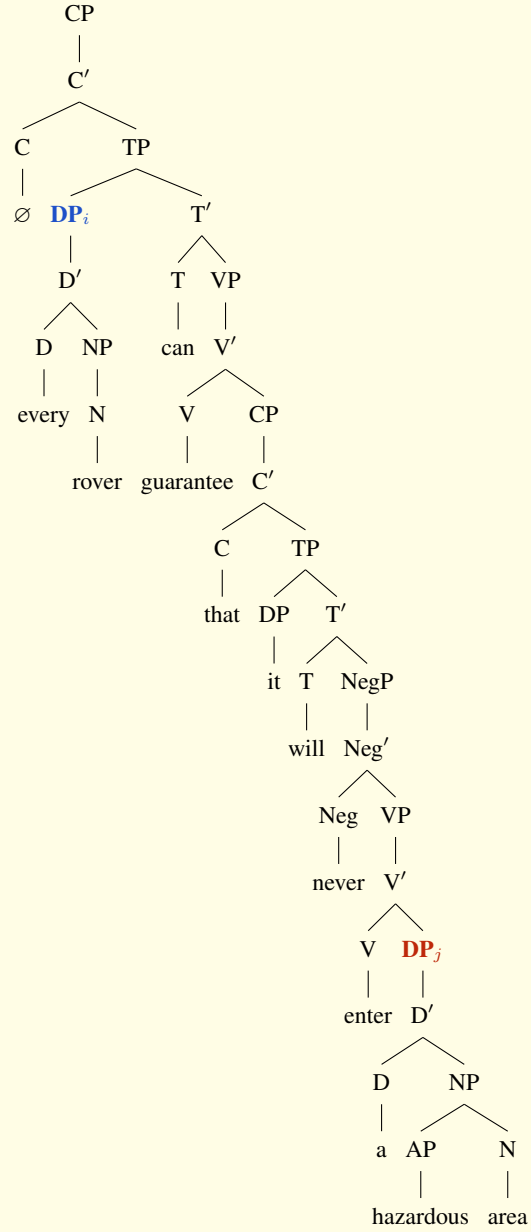

\subsection{Example 2: Surface-Scope Interpretation}

\begin{example}[Surface-scope interpretation: $\forall > \exists$]
Under the surface-scope interpretation, the universal quantifier takes wider scope than the existential quantifier. Informally, the reading is: "for every rover, there is a possibly different hazardous area such that the rover can guarantee that it will never enter it".
\end{example}

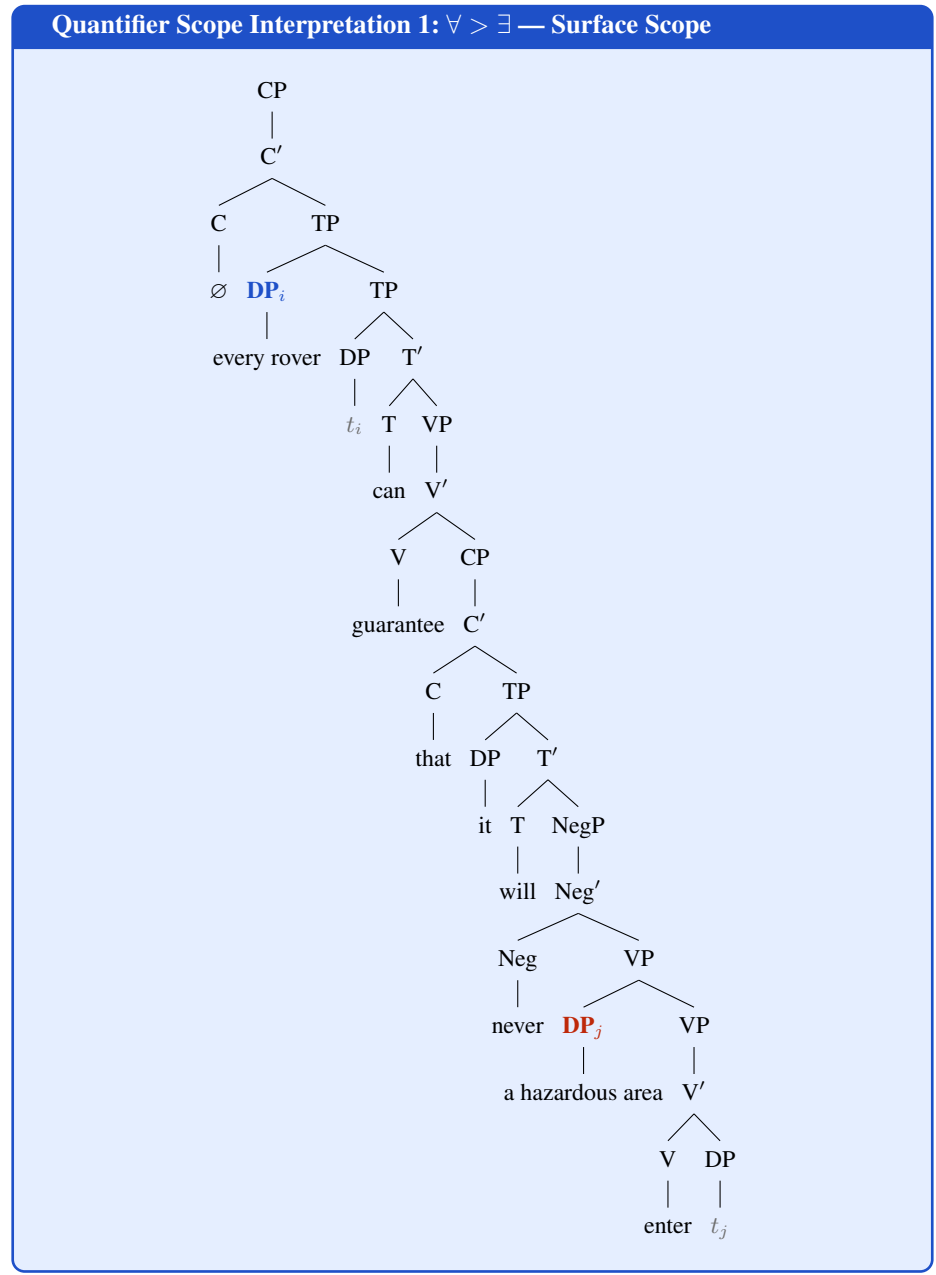
\begin{figure}[t]
\centering
\begin{tcolorbox}[
  enhanced,
  arc=4pt,
  boxrule=1pt,
  colback=BG1!70,
  colframe=QP1c,
  title={\bfseries Quantifier Scope Interpretation 1: $\forall > \exists$ --- Surface Scope},
  fonttitle=\normalsize
]
\centering
\begin{adjustbox}{max width=\linewidth}
\begin{forest} xt
[CP
  [C$'$
    [C [$\varnothing$]]
    [TP
      [\QPA{DP$_i$} [every rover]]
      [TP
        [DP [\Tr{i}]]
        [T$'$
          [T [can]]
          [VP
            [V$'$
              [V [guarantee]]
              [CP
                [C$'$
                  [C [that]]
                  [TP
                    [DP [it]]
                    [T$'$
                      [T [will]]
                      [NegP
                        [Neg$'$
                          [Neg [never]]
                          [VP
                            [\QPB{DP$_j$} [a hazardous area]]
                            [VP
                              [V$'$ [V [enter]] [DP [\Tr{j}]]]
                            ]
                          ]
                        ]
                      ]
                    ]
                  ]
                ]
              ]
            ]
          ]
        ]
      ]
    ]
  ]
]
\end{forest}
\end{adjustbox}
\end{tcolorbox}
\caption{Surface-scope interpretation, where the universal DP outscopes the indefinite DP.}
\label{fig:qsa-surface-scope}
\end{figure}

\subsection{Example 3: Inverse-Scope Interpretation}

\begin{example}[Inverse-scope interpretation: $\exists > \forall$]
Under the inverse-scope interpretation, the indefinite DP receives wider semantic scope than the universal DP. Informally, the reading is: "there is a hazardous area such that every rover can guarantee that it will never enter that same area".
\end{example}

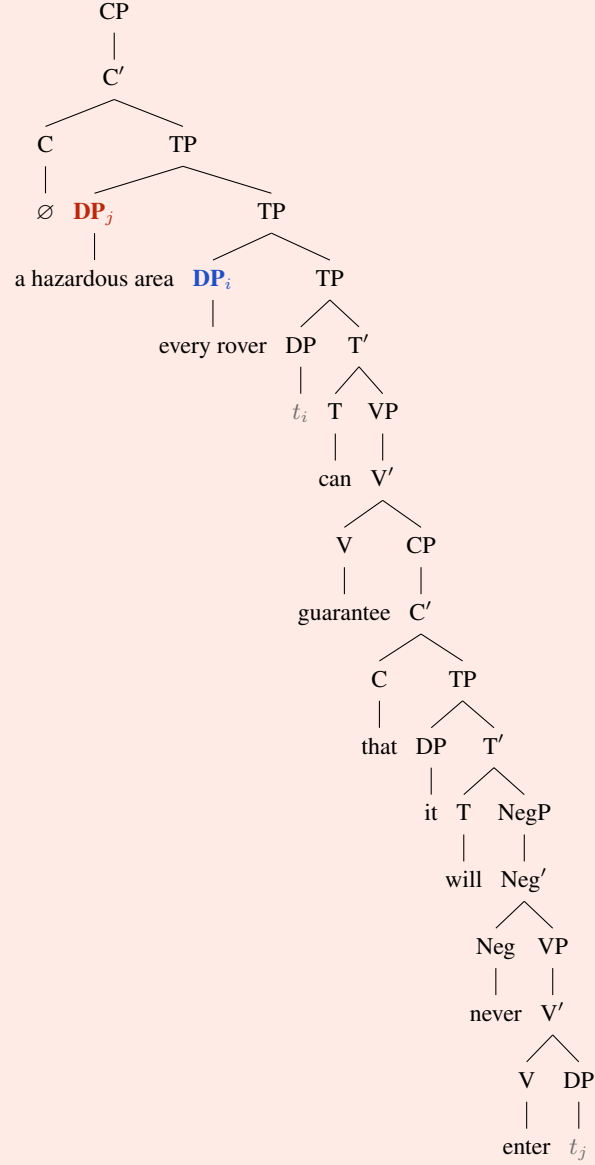
\begin{figure}[t]
\centering
\begin{tcolorbox}[
  enhanced,
  arc=4pt,
  boxrule=1pt,
  colback=BG2!70,
  colframe=QP2c,
  title={\bfseries Quantifier Scope Interpretation 2: $\exists > \forall$ --- Inverse Scope},
  fonttitle=\normalsize
]
\centering
\begin{adjustbox}{max width=\linewidth}
\begin{forest} xt
[CP
  [C$'$
    [C [$\varnothing$]]
    [TP
      [\QPB{DP$_j$} [a hazardous area]]
      [TP
        [\QPA{DP$_i$} [every rover]]
        [TP
          [DP [\Tr{i}]]
          [T$'$
            [T [can]]
            [VP
              [V$'$
                [V [guarantee]]
                [CP
                  [C$'$
                    [C [that]]
                    [TP
                      [DP [it]]
                      [T$'$
                        [T [will]]
                        [NegP
                          [Neg$'$
                            [Neg [never]]
                            [VP
                              [V$'$ [V [enter]] [DP [\Tr{j}]]]
                            ]
                          ]
                        ]
                      ]
                    ]
                  ]
                ]
              ]
            ]
          ]
         Ketch
        ]
      ]
    ]
  ]
]
\end{forest}
\end{adjustbox}
\end{tcolorbox}
\caption{Inverse-scope interpretation, where the indefinite DP receives wider semantic scope than the universal DP.}
\label{fig:qsa-inverse-scope}
\end{figure}

The tree in Figure~\ref{fig:qsa-inverse-scope} employs a scope-marking convention rather than a syntactic derivation. The indefinite DP "a hazardous area" is shown TP-adjoined to indicate its wide-scope existential interpretation, not to assert a literal movement operation. This representation is compatible with analyses in which indefinites receive exceptional scope via choice functions \cite{reinhart1997quantifier} or referentiality \cite{fodor1982referential}, without violating island constraints on QR. It should not be interpreted as a claim about syntactic derivation.

\subsection{Example 4: VP-Ellipsis under Sentential Coordination}

\begin{example}[VP-ellipsis]
Consider the sentence:
\begin{quote}
The captain can guarantee that at the next step possession will be recovered, and the midfielder can too.
\end{quote}
The second conjunct contains an elliptical VP. The phrase \emph{can too} must be interpreted by recovering the full VP from the first conjunct. Consequently, the ATL/ATL$^\ast$ translation must duplicate the strategic-temporal objective for the second agent.
\end{example}

\begin{figure}[t]
\centering
\begin{tcolorbox}[
  enhanced,
  arc=4pt,
  boxrule=1pt,
  colback=Hdc!5,
  colframe=Ellc,
  title={\bfseries VP-Ellipsis},
  fonttitle=\normalsize
]
\centering
\begin{adjustbox}{max width=\linewidth}
\begin{forest} xt
[CoordP
  [TP$_1$
    [DP [The captain]]
    [T$'$
      [T [can]]
      [VP$_i$
        [V$'$
          [V [guarantee]]
          [CP
            [C$'$
              [C [that]]
              [TP
                [PP [at the next step]]
                [TP
                  [DP [possession]]
                  [T$'$ [T [will]] [VP [be recovered]]]
                ]
              ]
            ]
          ]
        ]
      ]
    ]
  ]
  [Coord$'$
    [Coord [and]]
    [TP$_2$
      [DP [the midfielder]]
      [T$'$
        [T [can]]
        [VP [$e_i$]]
       Dean]
    ]
  ]
]
\end{forest}
\end{adjustbox}
\end{tcolorbox}
\caption{VP-ellipsis under sentential coordination. The elided VP in the second conjunct is represented as an indexed empty category $e_i$, co-indexed with the antecedent VP$_i$ in the first conjunct. This notation indicates identity of interpretation without commitment to the theoretical mechanism (deletion vs. pro-form).}
\label{fig:vp-ellipsis}
\end{figure}
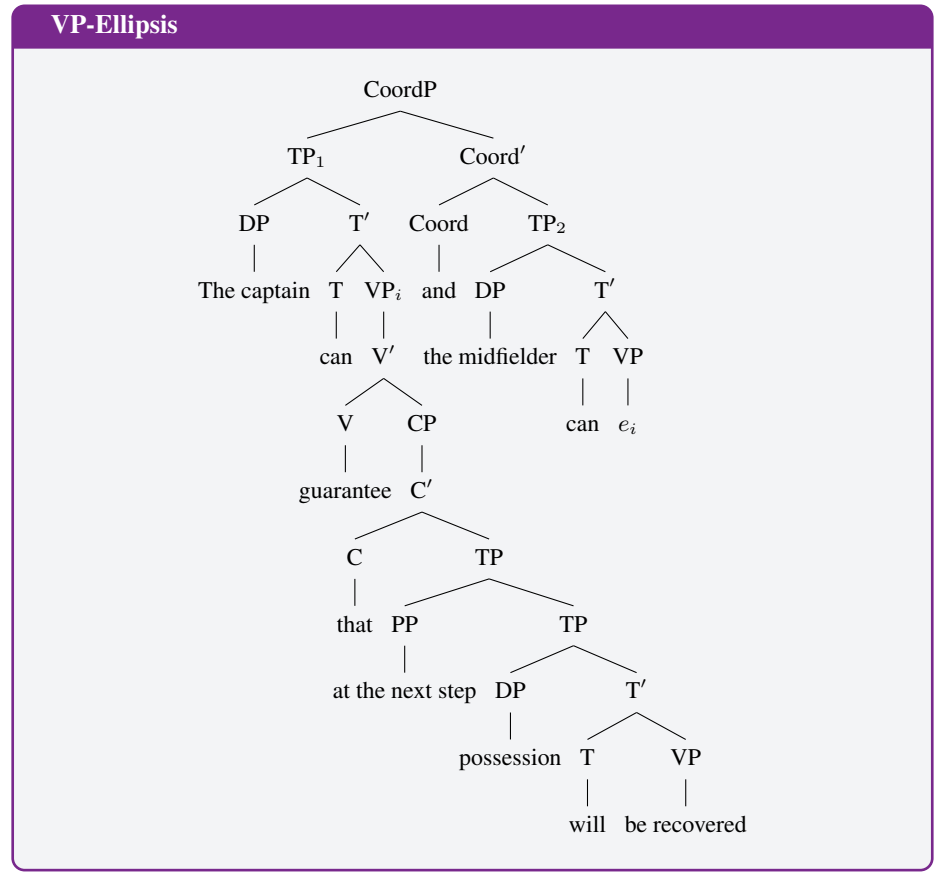

The corresponding translation reconstructs the elided material before formalization:
\[
\atl{Captain}X\,possession\_recovered
\;\wedge\;
\atl{Midfielder}X\,possession\_recovered.
\]
Thus, the example tests whether a model can recover omitted material and duplicate the full strategic-temporal objective, rather than translating only the overt surface string.

\subsection{Example 6: Right Dislocation}
 
\begin{example}[Right dislocation]
Consider the sentence:
\begin{quote}
  They can eventually ensure the game ends, the players.
\end{quote}
The NP \emph{the players} appears at the right periphery of the matrix clause,
outside its canonical argument position.
In conformity with X-bar theory's binary-branching requirement,
the dislocated element is represented as a right-adjunct:
the outer TP has exactly two daughters, the inner TP (the base clause)
and the right-adjoined $\mathrm{DP}_i$, so that no node carries more
than two immediate children.
The subject pronoun \emph{they} occupies the base-generated subject position
and is coreferential with the dislocated NP (co-indexed as $\mathrm{DP}_i$).
The adverb \emph{eventually} adjoins to VP, itself producing a binary VP-over-VP
adjunction structure.
\end{example}
 
\begin{figure}[t]
\centering
\begin{tcolorbox}[
  enhanced,
  arc=4pt,
  boxrule=1pt,
  colback=BG3!70,
  colframe=RDc,
  title={\bfseries Right Dislocation},
  fonttitle=\normalsize
]
\centering
\begin{adjustbox}{max width=\linewidth}
\begin{forest} xt
[TP
  [TP
    [\RDA{DP$_i$} [they]]
    [T$'$
      [T [can]]
      [VP
        [AdvP [Adv [eventually]]]
        [VP
          [V$'$
            [V [ensure]]
            [CP
              [C$'$
                [C [$\varnothing$]]
                [TP
                  [DP [D$'$ [D [the]] [NP [N [game]]]]]
                  [T$'$ 
                    [T [$\varnothing$]] 
                    [VP [V$'$ [V [ends]]]]
                  ]
                ]
              ]
            ]
          ]
        ]
      ]
    ]
  ]
  [\RDA{DP$_i$}\\{\scriptsize[right-adjoined]}
    [D$'$ [D [the]] [NP [N [players]]]]
  ]
]
\end{forest}
\end{adjustbox}
\end{tcolorbox}
\caption{Right dislocation via right-adjunction.
The outer TP has exactly two daughters: the inner TP (the base clause)
and the right-adjoined $\mathrm{DP}_i$ \emph{the players} (in \textcolor{RDc}{\bfseries green}).
The resumptive subject \emph{they} and the dislocated NP share index $i$.
The adverb \emph{eventually} adjoins to VP, yielding a VP-over-VP binary structure.}
\label{fig:right-dislocation}
\end{figure}
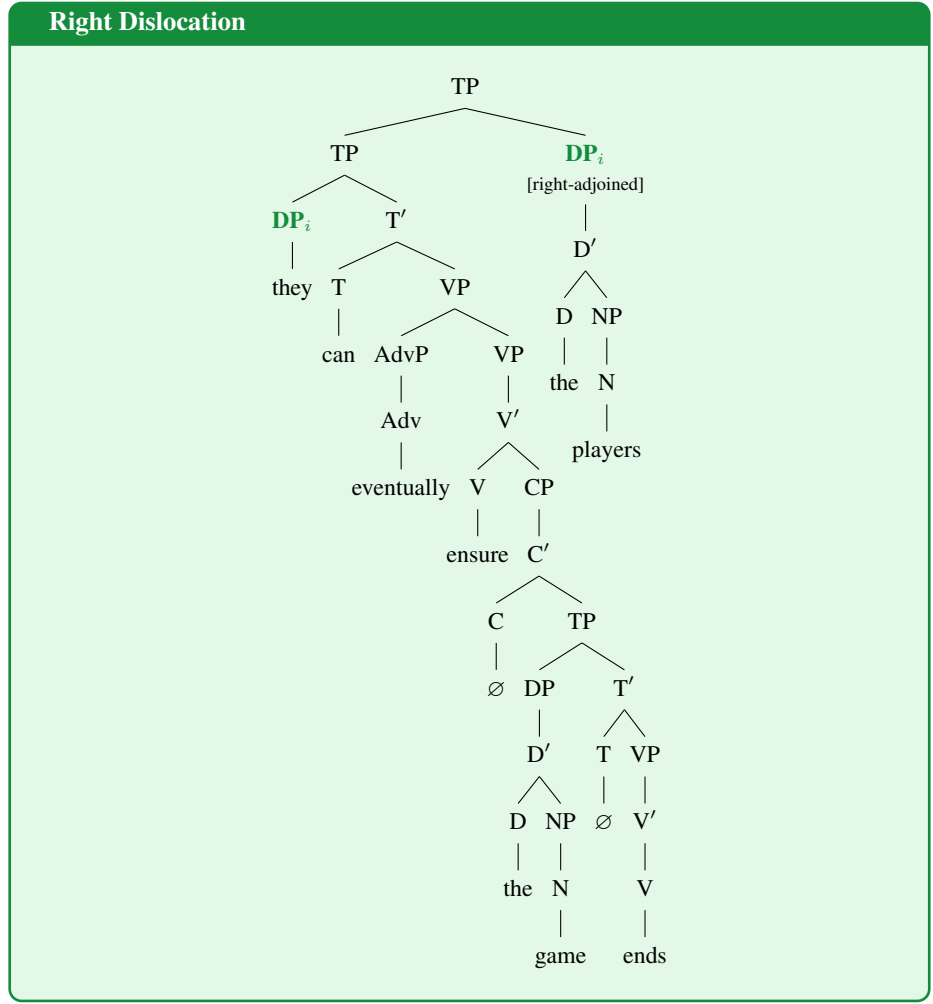
 
\subsection{Example 7: Left Dislocation}
 
\begin{example}[Left dislocation]
Consider the sentence:
\begin{quote}
  The vending machine, Mary can ensure it will work.
\end{quote}
The DP \emph{the vending machine} is displaced to the left periphery of the clause,
occupying Spec,TopP in the cartographic analysis of the left periphery
\cite{rizzi1997fine}. A resumptive pronoun \emph{it} fills the canonical argument
position within the embedded complement of \emph{ensure}, co-indexed
with the topic $\mathrm{DP}_i$. Binary branching is satisfied throughout:
TopP branches into Spec and Top$'$; Top$'$ branches into Top$^\circ$ and TP;
and so on down every projection.
\end{example}
 
\begin{figure}[t]
\centering
\begin{tcolorbox}[
  enhanced,
  arc=4pt,
  boxrule=1pt,
  colback=BG4!70,
  colframe=LDc,
  title={\bfseries Left Dislocation},
  fonttitle=\normalsize
]
\centering
\begin{adjustbox}{max width=\linewidth}
\begin{forest} xt
[TopP
  [\LDA{DP$_i$}
    [D$'$ [D [the]] [NP [N [vending\\machine]]]]
  ]
  [Top$'$
    [Top [$\varnothing$]]
    [TP
      [DP [Mary]]
      [T$'$
        [T [can]]
        [VP
          [V$'$
            [V [ensure]]
            [CP
              [C$'$
                [C [$\varnothing$]]
                [TP
                  [\LDA{DP$_i$} [it]]
                  [T$'$
                    [T [will]]
                    [VP [V$'$ [V [work]]]]
                  ]
                ]
              ]
            ]
          ]
        ]
      ]
    ]
  ]
]
\end{forest}
\end{adjustbox}
\end{tcolorbox}
\caption{Left dislocation: \emph{the vending machine} ($\mathrm{DP}_i$,
in \textcolor{LDc}{\bfseries amber}) occupies Spec,TopP in the left periphery.
The resumptive pronoun \emph{it} ($\mathrm{DP}_i$) within the embedded clause
is co-indexed with the topic. Top$^\circ$ is phonologically null ($\varnothing$).
All nodes branch at most binary.}
\label{fig:left-dislocation}
\end{figure}
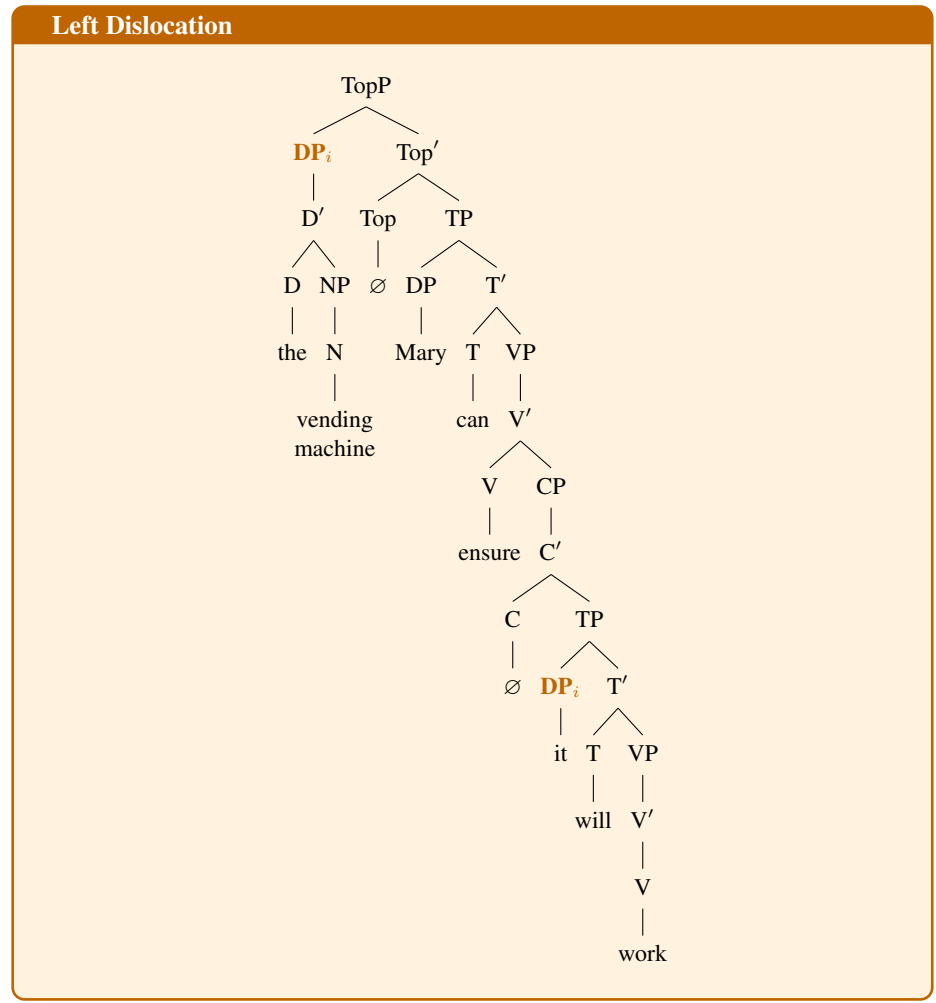
 
\subsection{Example 8: Right Node Raising}
 
\begin{example}[Right Node Raising]
Consider the sentence:
\begin{quote}
  Bob has, but Albert has not, a strategy to ensure that the vending machine
  is always operative.
\end{quote}
The DP \emph{a strategy to ensure that the vending machine is always operative}
is shared between two coordinated TPs.
In conformity with X-bar theory's binary-branching requirement,
the shared element is represented as a right-adjunct to CoordP:
the outer CoordP has exactly two daughters, the inner CoordP
(the coordinate structure proper) and the right-adjoined $\mathrm{DP}_j$,
so that no node carries more than two immediate children.
Each conjunct contains a gap in object position, marked by trace $t_j$.
To reflect the formal syntax of the second conjunct where the verb precedes negation (\emph{has not}), the verb \emph{has} is raised to T in both conjuncts, leaving a trace $t_v$ heading the VP. 
The negation \emph{not} heads a NegP below T$'$.
The internal structure of the complex shared DP is omitted for readability.
\end{example}
 
\begin{figure}[t]
\centering
\begin{tcolorbox}[
  enhanced,
  arc=4pt,
  boxrule=1pt,
  colback=BG5!70,
  colframe=RNRc,
  title={\bfseries Right Node Raising},
  fonttitle=\normalsize
]
\centering
\begin{adjustbox}{max width=\linewidth}
\begin{forest} xt
[CoordP
  [CoordP
    [TP$_1$
      [DP [Bob]]
      [T$'$
        [T [has]]
        [VP [V$'$ [V [$t_v$]] [DP [\Tr{j}]]]]
      ]
    ]
    [Coord$'$
      [Coord [but]]
      [TP$_2$
        [DP [Albert]]
        [T$'$
          [T [has]]
          [NegP
            [Neg$'$
              [Neg [not]]
              [VP [V$'$ [V [$t_v$]] [DP [\Tr{j}]]]]
            ]
          ]
        ]
      ]
    ]
  ]
  [\RNRA{DP$_j$}\\{\scriptsize[right-adjoined / shared]}
    [D$'$ [D [a]] [NP [N [strategy\\to ensure $\ldots$]]]]
  ]
]
\end{forest}
\end{adjustbox}
\end{tcolorbox}
\caption{Right Node Raising via right-adjunction. The shared DP is represented as right-adjoined to CoordP to maintain binary branching while ensuring the ATL translation associates the objective with both conjuncts. This is a representational convention for the syntax-semantics interface, not a syntactic derivation. The traces $t_j$ in object position mark interpretive gaps; no movement operation is claimed.}
\label{fig:rnr}
\end{figure}
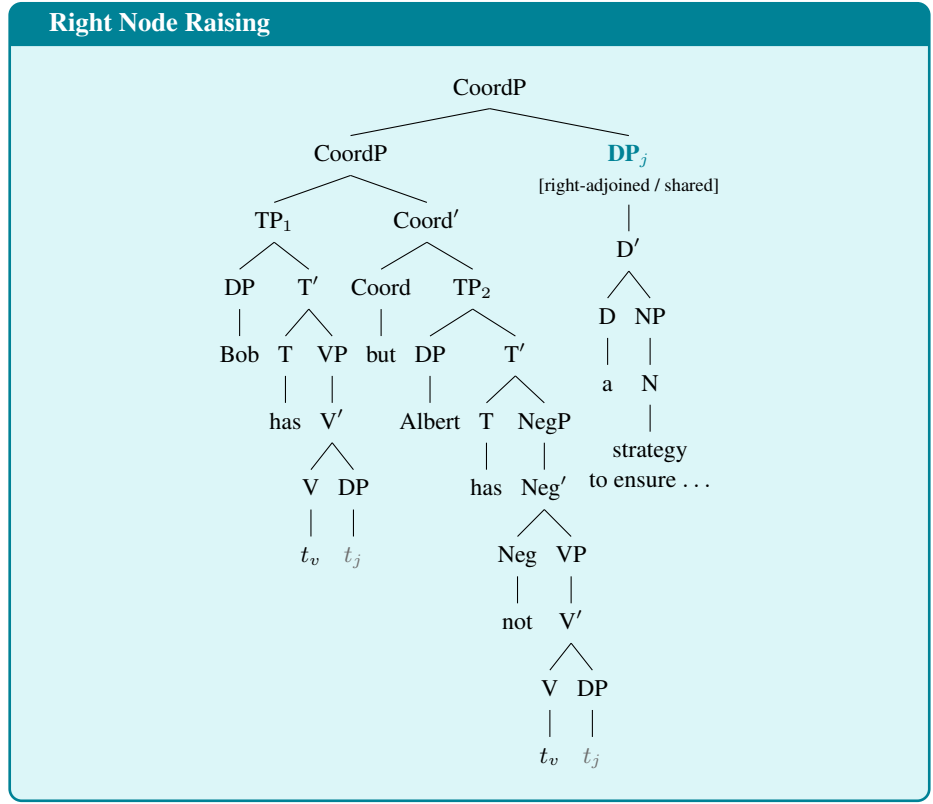

\section{Model-Checker Integration Interface}

\begin{figure}[h]
    \centering
    \includegraphics[width=\linewidth]{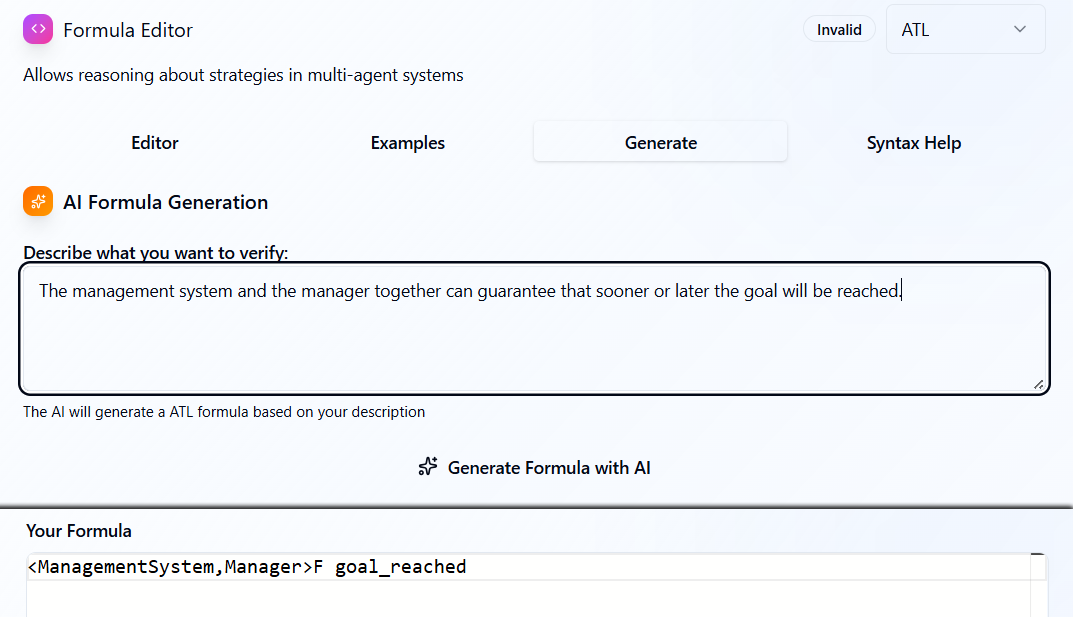}
    \caption{Natural-language to ATL$^\star$ translation surfaced inside the \textsc{genVITAMIN} interface.}
    \label{fig:vitamin}
\end{figure}

To make the end-to-end workflow concrete, Figure~\ref{fig:vitamin} shows the natural-language-to-ATL$^\ast$ translation surfaced directly inside the \textsc{VITAMIN} model checker. A user enters a strategic requirement in natural language; the translation service returns a well-formed ATL/ATL$^\ast$ formula that is parsed by the same front-end used for verification, so that only syntactically valid specifications enter the model-checking phase.

\section{Relevance to the Dataset}

These examples illustrate why ambiguity-aware NL-to-ATL/ATL$^\ast$ translation cannot be reduced to direct keyword substitution. Quantifier scope ambiguity may require multiple admissible formal outputs, while VP-ellipsis requires reconstruction of omitted material before translation. Similar considerations apply to right dislocation, left dislocation, and Right Node Raising. In all cases, the resulting formula must preserve the intended scope relations among strategic modalities, temporal operators, Boolean connectives, and atomic propositions.

\section{Translation-and-Evaluation Algorithm}
\label{app:algorithm}

Algorithm~\ref{alg:pipeline} details the end-to-end translation-and-evaluation loop that the \texttt{nl2atl} architecture (Figure~1 in the main text) realizes: it draws a fixed stratified split, fine-tunes each open-weight model on the training split (skipped for the API baselines), generates a prediction for every test requirement under greedy decoding, and scores each prediction with the two-tier exact-match-then-judge protocol before aggregating the per-prediction verdicts over seeds.

\begin{algorithm}[t]
\caption{Translation and Evaluation}
\label{alg:pipeline}
\begin{algorithmic}[1]
\REQUIRE Dataset $\mathcal{D}$, model config $M$, training seeds $S$
\ENSURE Raw predictions $\mathcal{P}$, per-prediction verdicts $\mathcal{V}$, and aggregated accuracy
\STATE $(\mathcal{D}_{\mathrm{tr}}, \mathcal{D}_{\mathrm{te}}) \gets \textsc{StratifiedSplit}(\mathcal{D})$ \COMMENT{fixed canonical split; few-shot exemplars held out}
\STATE Instantiate the inference backend for $M$ via the Model Abstraction Layer
\STATE $\mathcal{P} \gets \emptyset$;\quad $\mathcal{V} \gets \emptyset$ \COMMENT{raw predictions; per-prediction verdicts}
\FORALL{training seed $s \in S$}
    \STATE $M_s \gets \textsc{FineTune}(M, \mathcal{D}_{\mathrm{tr}}, s)$ with LoRA \COMMENT{skipped for API baselines}
    \FORALL{$r \in \mathcal{D}_{\mathrm{te}}$}
        \STATE $P \gets \textsc{BuildPrompt}(r, M)$ \COMMENT{system prompt $+$ optional few-shot}
        \STATE $\hat{\phi} \gets \textsc{Infer}(P, M_s)$ \COMMENT{greedy decoding; local or API backend}
        \STATE $\mathcal{P} \gets \mathcal{P} \cup \{(r, \hat{\phi}, s, \text{metadata})\}$
    \ENDFOR
\ENDFOR
\FORALL{$(r, \hat{\phi}, s) \in \mathcal{P}$ with gold set $\Phi_{\mathrm{ref}}$}
    \IF{$\textsc{ExactMatch}(\hat{\phi}, \Phi_{\mathrm{ref}})$}
        \STATE $\mathit{correct} \gets \TRUE$
    \ELSE
        \STATE $\mathit{correct} \gets \textsc{LLMJudge}(\hat{\phi}, \Phi_{\mathrm{ref}})$
    \ENDIF
    \STATE $\mathcal{V} \gets \mathcal{V} \cup \{(r, s, \mathit{correct})\}$ \COMMENT{store verdict for aggregation}
\ENDFOR
\STATE $\mathit{accuracy} \gets \textsc{Aggregate}(\mathcal{V}, S)$ \COMMENT{mean over seeds with dispersion}
\RETURN $\mathcal{P}$, $\mathcal{V}$, and $\mathit{accuracy}$
\end{algorithmic}
\end{algorithm}

\section{Experimental Configuration}
\label{app:config}

\subsection{Computing Infrastructure and Software}
All local fine-tuning and inference were run on Linux compute nodes equipped with NVIDIA A100 GPUs, dispatched as independent jobs through a SLURM-managed cluster. The proprietary baselines and the LLM judges were accessed through the Azure OpenAI service. The framework is implemented in Python ($\geq 3.10$) and released under the MIT license. The experiments build, among others, on PyTorch~2.10, Hugging Face Transformers~5.0, PEFT~0.18 (low-rank adaptation), \texttt{bitsandbytes}~0.49 ($4$-bit quantization), TRL~0.27 (supervised fine-tuning), Accelerate~1.12, Datasets~4.5, and scikit-learn~1.7. The released repository pins an exact version for every dependency, and the per-node CPU and memory specifications are recorded in the accompanying environment manifest.

\subsection{Model Versions and Endpoints}
{\sloppy\setlength{\emergencystretch}{3em}
Every open-weight base checkpoint is pinned to an exact Hugging Face commit revision (Table~\ref{tab:modelversions}), and the released configuration records these revisions so that fine-tuning and inference reproduce the same weights. The proprietary baselines and four of the six LLM judges are Azure OpenAI deployments, queried with API version \texttt{2024-08-01-preview} at temperature~$0$ during June~2026. The generator deployments resolved to the dated snapshots \texttt{gpt-4.1-2025-04-14} and \texttt{gpt-5.4-2026-03-05}, and these two models also serve as judges; the other two Azure judges resolved to \texttt{gpt-5.2-2025-12-11} (\textsc{GPT-5.2}) and the deployed model version \texttt{DeepSeek-V3.2}. The remaining two judges are open-weight and run locally at $4$-bit precision like the open-weight generators: \textsc{Gemma-2-27B} (\texttt{google/gemma-2-27b-it}, revision \texttt{aaf20e6b}) and \textsc{Llama-3.3-70B} (\texttt{meta-llama/Llama-3.3-70B-Instruct}, revision \texttt{6f6073b4}).\par}

\begin{table}[h]
\centering
\small
\begin{tabular}{@{}lll@{}}
\toprule
\textbf{Model} & \textbf{Hugging Face checkpoint} & \textbf{Revision} \\
\midrule
qwen-3b       & Qwen/Qwen2.5-3B-Instruct           & \texttt{aa8e7253} \\
phi3          & microsoft/Phi-3.5-mini-instruct    & \texttt{2fe19245} \\
qwen-coder-7b & Qwen/Qwen2.5-Coder-7B-Instruct     & \texttt{c03e6d35} \\
mistral-7b    & mistralai/Mistral-7B-Instruct-v0.3 & \texttt{c170c708} \\
\bottomrule
\end{tabular}
\caption{Pinned open-weight base checkpoints and their exact Hugging Face commit revisions (short hashes). The proprietary generators (\textsc{gpt-4.1}, \textsc{gpt-5.4}) (which also serve as two of the six LLM judges) and the additional Azure judges (\textsc{DeepSeek-V3.2}, \textsc{GPT-5.2}) are Azure OpenAI deployments queried at API version \texttt{2024-08-01-preview}; these resolved to snapshots \texttt{gpt-4.1-2025-04-14}, \texttt{gpt-5.4-2026-03-05}, and \texttt{gpt-5.2-2025-12-11}, while \textsc{DeepSeek-V3.2} is identified by its model version. The other two judges, the open-weight \textsc{Gemma-2-27B} (\texttt{google/gemma-2-27b-it}) and \textsc{Llama-3.3-70B} (\texttt{meta-llama/Llama-3.3-70B-Instruct}), are run locally like the open-weight generators above.}
\label{tab:modelversions}
\end{table}

\subsection{Data Splits and Seeds}
The gold dataset is partitioned into stratified training/validation/test sets in a $70/10/20$ ratio, stratified on formula structure (single- versus multi-reading items). The split seed is decoupled from the training seed and fixed at $42$ for the canonical split used for all headline numbers; each fine-tuned configuration is then repeated over three training seeds ($42$, $43$, $44$), and accuracy is reported as the seed mean with a $95\%$ confidence interval. These intervals quantify sensitivity to the training seed at the \emph{fixed} canonical test split and therefore do not capture test-set sampling variance; for a binomial proportion at the observed accuracies, that component is on the order of $\pm0.06$ over the $218$ test items, so the seed intervals understate total uncertainty. The orchestrator additionally supports stratified $k$-fold cross-validation with shared folds, which estimates split-induced variance directly; we report the canonical single-split numbers as the headline and leave a full $k$-fold sweep across all configurations (which multiplies fine-tuning cost by the number of folds) to future runs. The curated few-shot exemplars are held out of every split so that no prompting example can leak into evaluation. Decoding is deterministic (greedy, with sampling disabled) and emits up to $256$ new tokens for local models; Azure calls are issued at temperature~$0$. Only the training split is augmented: each training instance is duplicated once with a templated paraphrase of its natural-language input - a single synonym substitution drawn from a fixed list of temporal and strategic phrasings (for example, ``sooner or later''\,$\rightarrow$\,``eventually'' or ``can guarantee that''\,$\rightarrow$\,``can ensure that'') - while its gold formula is left unchanged. With an augmentation factor of two this doubles the training data (the original instance plus one paraphrase); the validation and test splits are used verbatim, so augmentation cannot introduce train/test leakage.

\subsection{Training and LoRA Hyperparameters}
All open-weight models are fine-tuned with low-rank adapters (LoRA) for $8$ epochs using the paged $8$-bit AdamW optimizer, a cosine schedule with a $0.1$ warmup ratio, peak learning rate $1\times10^{-4}$, weight decay $0.01$, gradient clipping at $0.3$, \texttt{bf16} precision, and gradient checkpointing; the maximum sequence length is $1536$ tokens. Table~\ref{tab:hparams} lists the per-model adapter and batching settings. LoRA adapters are applied to the attention projections and the MLP projections of each architecture, and every base checkpoint is pinned to an exact revision in the released configuration. The proprietary baselines (\textsc{gpt-4.1}, \textsc{gpt-5.4}) are API-only and are not fine-tuned: their weights are closed, and the modest in-domain data available is insufficient to fine-tune models at their scale reliably, so they are evaluated zero- and few-shot.

\begin{table}[h]
\centering
\small
\begin{tabular}{@{}lccccc@{}}
\toprule
\textbf{Model} & \textbf{4-bit} & \textbf{LoRA }$r$ & \textbf{LoRA }$\alpha$ & \textbf{Dropout} & \textbf{Batch}$\,\times\,$\textbf{Acc.} \\
\midrule
mistral-7b     & yes & 32 & 64  & 0.05 & $2\times16$ \\
qwen-3b        & no  & 64 & 128 & 0.05 & $8\times4$  \\
phi3           & no  & 32 & 64  & 0.05 & $6\times6$  \\
qwen-coder-7b  & yes & 64 & 128 & 0.05 & $4\times8$  \\
\bottomrule
\end{tabular}
\caption{Per-model fine-tuning configuration. All models share the training schedule described in the text and a maximum sequence length of $1536$ tokens. ``Batch\,$\times$\,Acc.'' denotes the per-device batch size times the gradient-accumulation steps.}
\label{tab:hparams}
\end{table}

\section{Additional Quantitative Results}
\label{app:results}

\begin{figure}[t]
    \centering
    \includegraphics[width=0.7\linewidth]{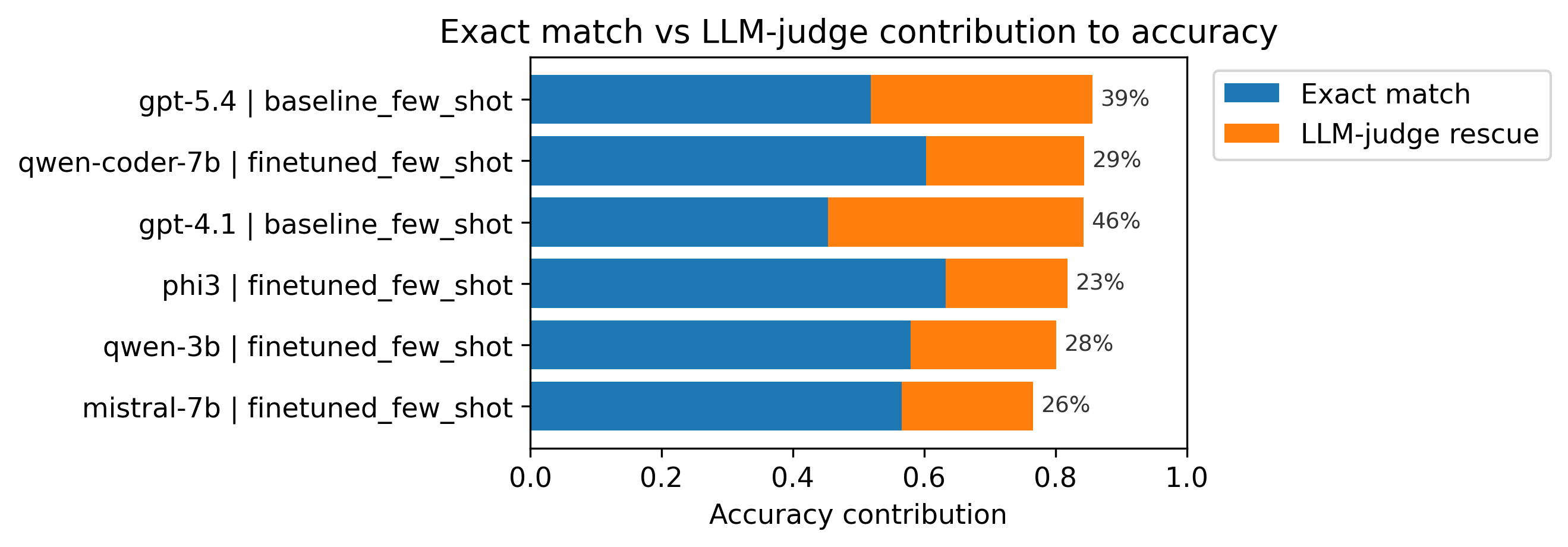}
    \caption{\textbf{Accuracy decomposition (Llama-3.3-70B judge).} Semantic accuracy under the headline judge split into the deterministic exact-match floor (blue) and the additional fraction recovered by the LLM judge (orange), for each headline system (proprietary few-shot baselines; open-weight fine-tuned few-shot). The judge-recovered share (annotated) is largest for the proprietary baselines (up to $46\%$ for \textsc{gpt-4.1} few-shot) and smaller for the fine-tuned open-weight systems ($23$--$29\%$), whose outputs more often match the reference surface form. Exact match alone would substantially understate the quality of every system.}
    \label{fig:decomposition}
\end{figure}

\begin{figure}[t]
    \centering
    \includegraphics[width=0.7\linewidth]{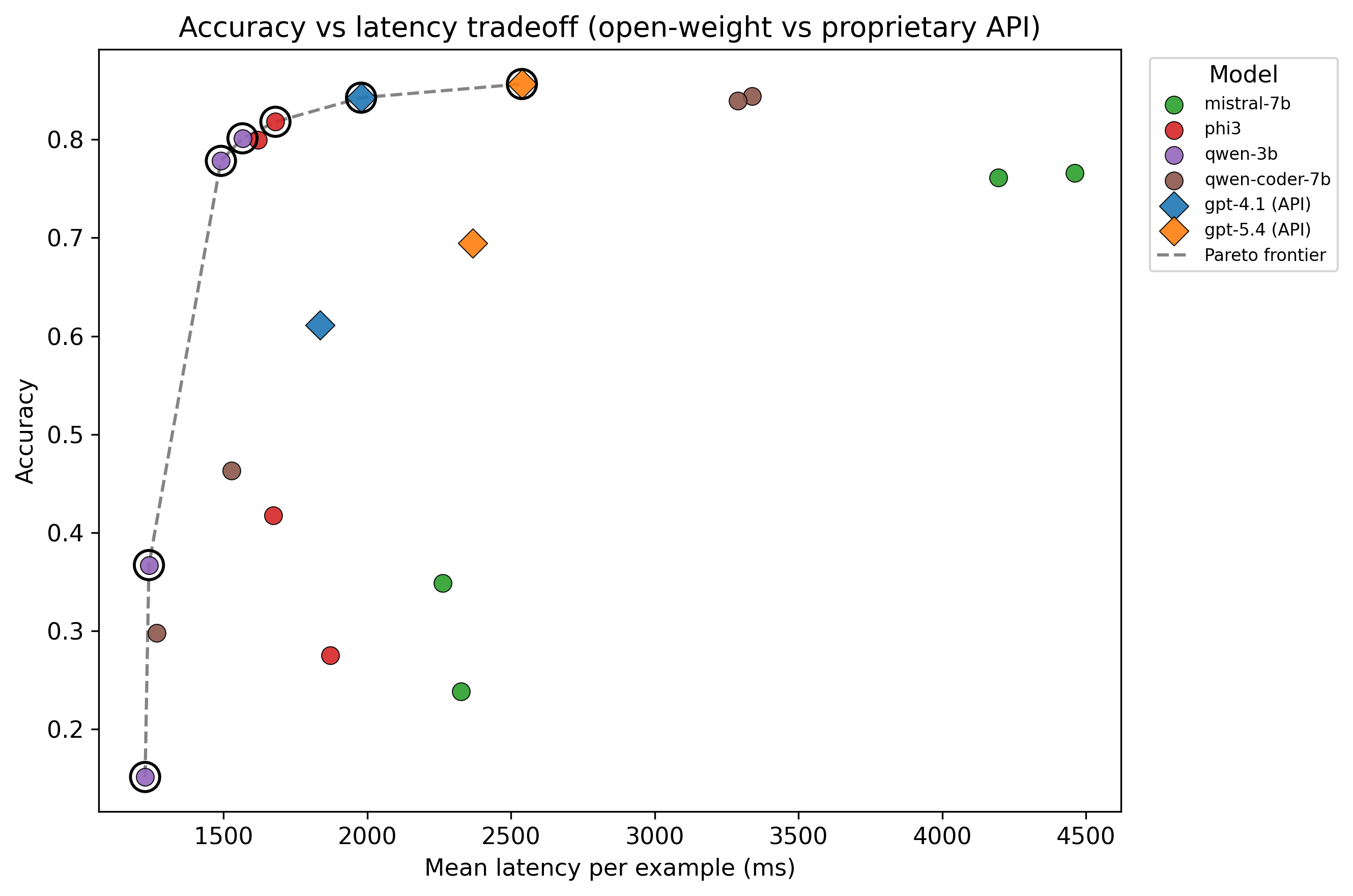}
    \caption{Semantic accuracy (under the headline Llama-3.3-70B judge) versus mean inference latency. Open-weight models are circles, proprietary API baselines diamonds; the dashed line is the Pareto frontier, which under the human-aligned judge is \emph{shared}: the few-shot proprietary baselines (\textsc{gpt-5.4}, \textsc{gpt-4.1}) occupy the high-accuracy end and the fast fine-tuned open-weight systems (\textsc{phi3}, \textsc{qwen-3b}) the low-latency end, while the most accurate open-weight system (\textsc{qwen-coder-7b}) sits just off the frontier, dominated by \textsc{gpt-5.4}. API latency is network-timed and only indicative across the API/local boundary; accuracy is comparable throughout.}
    \label{fig:accuracy_cost_tradeoff}
\end{figure}

This section reports the robustness check and fine-grained breakdowns referenced from the experiments in the main paper: the conservative six-judge accuracy (Table~\ref{tab:accuracy_sixjudge}), the decomposition of judged accuracy into its exact-match floor and judge-recovered fraction (Figure~\ref{fig:decomposition}), accuracy split by ambiguity type (Table~\ref{tab:qsa}), and the accuracy--latency trade-off (Figure~\ref{fig:accuracy_cost_tradeoff}).

\begin{table}[t]
\centering
\small
\setlength{\tabcolsep}{4pt}
\begin{tabular}{@{}llcccc@{}}
\toprule
& & \multicolumn{2}{c}{\textbf{Baseline}} & \multicolumn{2}{c}{\textbf{Fine-tuned}} \\
\cmidrule(lr){3-4}\cmidrule(lr){5-6}
\textbf{Model} & \textbf{Size} & ZS & FS & ZS & FS \\
\midrule
\multicolumn{6}{@{}l}{\textit{Proprietary (Azure API)}}\\
gpt-4.1        & API  & 0.438 & 0.623 & --    & --    \\
gpt-5.4        & API  & \textbf{0.445} & \textbf{0.681} & --    & --    \\
\midrule
\multicolumn{6}{@{}l}{\textit{Open-weight (local)}}\\
mistral-7b     & 7B   & 0.115 & 0.223 & 0.629\,{\scriptsize$\pm$.04} & 0.665\,{\scriptsize$\pm$.05} \\
qwen-3b        & 3B   & 0.070 & 0.265 & 0.667\,{\scriptsize$\pm$.04} & 0.692\,{\scriptsize$\pm$.02} \\
phi3           & 3.8B & 0.147 & 0.293 & 0.707\,{\scriptsize$\pm$.02} & 0.728\,{\scriptsize$\pm$.02} \\
qwen-coder-7b  & 7B   & 0.184 & 0.356 & \textbf{0.716}\,{\scriptsize$\pm$.02} & \textbf{0.731}\,{\scriptsize$\pm$.02} \\
\bottomrule
\end{tabular}
\caption{\textbf{Robustness check (six-judge mean).} Semantic accuracy averaged over all \emph{six} LLM judges and over seeds, a complement to the human-aligned primary results in the main paper (Table~1). Folding the stricter judges into the mean (down to GPT-5.2 at Cohen's $\kappa=0.16$ against humans) lowers every score and slightly reorders the top: the fine-tuned \textsc{qwen-coder-7b} ($0.731$) and \textsc{phi3} ($0.728$) edge above the strongest few-shot proprietary baseline (\textsc{gpt-5.4}, $0.681$) by a small margin that is significant on shared items (\textsc{qwen-coder-7b} $\Delta=0.052$, paired bootstrap $95\%$ CI $[0.004,0.102]$, randomization $p=0.03$). Under the single most human-aligned judge of Table~1 (which is also the most lenient) this margin vanishes to a tie ($\Delta=-0.009$, $p=0.71$); it reflects the strict judges' documented over-rejection of the proprietary baselines' faithful input-grounded paraphrases, not a true quality gap, so the honest reading is parity. Fine-tuned cells show the $95\%$ confidence interval over three training seeds; the local baselines are single greedy runs (deterministic, hence exactly reproducible), whereas the proprietary API baselines, though queried at temperature~$0$, are not bit-for-bit deterministic, so their single-run point estimates carry a run-to-run variance that we do not quantify. ZS/FS denote zero-/few-shot prompting; best per column in bold.}
\label{tab:accuracy_sixjudge}
\end{table}

\begin{table}[t]
\centering
\small
\begin{tabular}{@{}lcc@{}}
\toprule
\textbf{System} & \textbf{Single-reading} & \textbf{QSA} \\
 & ($n{=}187$) & ($n{=}31$) \\
\midrule
\textsc{gpt-4.1} (fs)          & 0.88 & 0.58 \\
\textsc{gpt-5.4} (fs)          & 0.88 & 0.68 \\
\midrule
\textsc{mistral-7b} (ft+fs)    & 0.85 & 0.27 \\
\textsc{qwen-3b} (ft+fs)       & 0.88 & 0.33 \\
\textsc{phi3} (ft+fs)          & 0.87 & 0.49 \\
\textsc{qwen-coder-7b} (ft+fs) & 0.88 & 0.63 \\
\bottomrule
\end{tabular}
\caption{Accuracy on the $187$ single-reading versus the $31$ multi-reading (quantifier-scope ambiguity, QSA) test items, under the headline Llama-3.3-70B judge, for the headline systems (proprietary few-shot baselines; fine-tuned few-shot open-weight models; ft+fs\,=\,fine-tuned\,+\,few-shot). Multi-reading items are the rare, hardest cases that require emitting both the distributive and the collective reading. Single-reading accuracy is uniformly high ($0.85$--$0.88$), but every system drops sharply on the QSA slice, where no model class dominates: the strongest few-shot proprietary baseline (\textsc{gpt-5.4}, $0.68$) and the best fine-tuned open-weight model (\textsc{qwen-coder-7b}, $0.63$) handle it comparably, while the smaller fine-tuned models lag ($0.27$--$0.33$). The QSA slice has only $31$ items (a binomial $95\%$ confidence interval is $\approx\pm0.17$), so it supports this qualitative contrast but not a fine ranking. Under the strictest judges the proprietary baselines instead score near zero on this slice; that collapse is an artifact of those judges rejecting their input-grounded predicate paraphrases (our judge-reliability analysis in the main paper), not a failure to emit both readings. This breakdown reuses the existing judged verdicts; no additional labeling was performed.}
\label{tab:qsa}
\end{table}

\section{Prompting and Few-Shot Exemplars}

All systems share a fixed prompting setup, identical across models and prompting conditions; the verbatim templates ship with the released code. We summarize their content here.

\subsection{Translation Prompt}
The system prompt casts the model as an ATL/ATL$^\ast$ expert and requires it to return \emph{only} formula text. It fixes:
\begin{itemize}[leftmargin=1.2em,itemsep=2pt]
  \item \textbf{Syntax.} The coalition modality \texttt{<<A>>} or \texttt{<<A,B>>}; the temporal operators \texttt{X}, \texttt{F}, \texttt{G} (unary) and \texttt{U} (binary, written \texttt{p U q}); the Boolean operators \texttt{!}, \texttt{\&\&}, \texttt{||}, \texttt{->}; \textsc{PascalCase} agent and coalition names; and \texttt{snake\_case} atomic propositions.
  \item \textbf{Scope.} The strategic operator scopes over the whole formula that follows it and is kept separate from the temporal operator it governs, e.g.\ \texttt{<<Machine>>G(paid -> ticket\_printed)}, not \texttt{<<Machine>>(G paid -> ticket\_printed)}. Inability is expressed by negating the strategic operator (\texttt{!<<Y>>F goal}), not the objective.
  \item \textbf{Targeted ambiguities.} VP ellipsis: repeat the recovered formula for the second agent. Right Node Raising: attach the shared right-peripheral objective to both conjuncts. Quantifier-scope ambiguity: output \emph{all} admissible readings, one per line, never fusing two readings into one formula; a distributive reading ascribes the ability to each agent separately and a collective reading to the coalition jointly.
  \item \textbf{Output discipline.} Emit only the formula(s), one per line, with no explanations, Markdown, labels, or trailing prose.
\end{itemize}

\subsection{Few-Shot Exemplars}
In the few-shot condition a fixed pool of seven curated exemplars is prepended to the prompt; these inputs are held out of every train/validation/test split, so they never leak into evaluation. Each exemplar targets a distinct phenomenon:
{\small\sloppy
\begin{itemize}[leftmargin=1.2em,itemsep=3pt]
  \item \textbf{Collective ability (right dislocation).} ``They can guarantee that at the next step the alarm will be sent, the surveillance system and the operator.''\\ $\Rightarrow$ \texttt{<<System,Operator>>X alarm\_sent}
  \item \textbf{Left dislocation.} ``The gate, the machine can guarantee that it will open at the next step.''\\ $\Rightarrow$ \texttt{<<Machine>>X gate\_open}
  \item \textbf{VP ellipsis.} ``Robot number 1 has a strategy to ensure that eventually position 3 holds, and robot number 2 does too.''\\ $\Rightarrow$ \texttt{<<Robot1>>F pos3 \&\& <<Robot2>>F pos3}
  \item \textbf{Quantifier-scope ambiguity (two required readings).} ``Every robot can guarantee that it will eventually reach a safe spot.''\\ $\Rightarrow$ \texttt{<<Robot1>>F at\_safe\_spot\_1 \&\& <<Robot2>>F at\_safe\_spot\_2 \&\& <<Robot3>>F at\_safe\_spot\_3}\\ and \texttt{<<Robot1,Robot2,Robot3>>F at\_safe\_spot}
  \item \textbf{Ability asymmetry.} ``The diplomatic cable system can, but the encryption gateway cannot, guarantee that classified cables will never be routed publicly.''\\ $\Rightarrow$ \texttt{<<DiplomaticCableSystem>>G !classified\_cables\_routed\_publicly \&\& !<<EncryptionGateway>>G !classified\_cables\_routed\_publicly}
  \item \textbf{Simple eventuality.} ``The user can guarantee that sooner or later the ticket will be printed.''\\ $\Rightarrow$ \texttt{<<User>>F ticket\_printed}
  \item \textbf{Nested, literary input.} ``If we do not wish to fight, we can prevent the enemy from engaging us\ldots'' (adapted from Sun Tzu).\\ $\Rightarrow$ \texttt{<<We>>(!wish\_to\_fight -> F (throw\_something\_odd\_in\_his\_way \&\& G !enemy\_engages\_us))}
\end{itemize}
}
With the count left unset, every exemplar is shown in a fixed order; a smaller count selects a reproducible random subset and is used only for ablations.

\subsection{LLM-Judge Prompt}
All six judges (DeepSeek-V3.2, GPT-4.1, GPT-5.2, GPT-5.4, Gemma-2-27B, and Llama-3.3-70B) use a single fixed prompt (version~v1.4) that casts the model as an adjudicator of ATL/ATL$^\ast$ \emph{faithfulness} beyond exact string matching. It restates the syntax conventions above and applies an explicit rubric.
\begin{itemize}[leftmargin=1.2em,itemsep=2pt]
  \item \textbf{Accept:} harmless whitespace and redundant parentheses; commutative reordering of \texttt{\&\&} or \texttt{||} when the same operands stay under the same strategic, temporal, and Boolean scope; and renamed predicates or agents only when they are clear aliases grounded in the input.
  \item \textbf{Any}\texttt{p -> q} as \texttt{!p || q}), contraposition, De Morgan, double-negation, idempotence, distributivity, or biconditional rewrites; any change in a temporal operator or its replacement by a temporal-logic equivalent (e.g.\ \texttt{F p} as \texttt{true U p}, \texttt{G p} as \texttt{!F !p}); any change of coalition, including distributive versus collective ability; any change of temporal or strategic scope, implication direction, or polarity; turning a conjunction into a disjunction or vise versa; and any omitted or extraneous condition. For ambiguous items it rejects predictions that return only one of the jointly required readings or that collapse them into a single conjunction.
  \item \textbf{Output and robustness.} The judge returns a single machine-parseable JSON object, \texttt{\{"correct": "yes" | "no", "reasoning": "..."\}}, and is calibrated with eight worked accept/reject examples (covering parenthesization, commutativity, operator rewrites, QSA multiplicity, and temporal-operator and coalition errors). To resist prompt injection, it is instructed to treat the input, gold output(s), and prediction as data and to ignore any instructions embedded inside them.
\end{itemize}
Verdicts are cached per judge identity, so a judge is never queried twice on an identical (input, gold, prediction) triple, while the judges remain independent.

\section{Human-Annotation Disagreements and Their Adjudication}
\label{appx:disagreements}

Of the $599$ audited predictions, the two expert annotators initially diverged on three. In all three the stricter annotator judged the prediction incorrect while the second judged it correct. The annotators then deliberated and reached agreement on two of the three (D2 and D3 below), folding those labels back into the reference set; the third (D1) resisted consensus and is retained here as a genuine unresolved disagreement rather than forced to a label. Both reconciled cases were settled in favour of the \emph{permissive} reading (the annotators agreed the prediction \emph{was} faithful) even though the automated judges had almost all called these two predictions incorrect (unanimously across the six judges for D2, and four-of-six for D3, where \textsc{gpt-4.1} and the human-aligned \textsc{Llama-3.3-70B} had sided with the permissive annotator). On these borderline items the human audit therefore overturns an over-strict tendency the judges share, rather than merely confirming them; even the most human-aligned judge, \textsc{Llama-3.3-70B}, shared this over-strict reading on D2, though it had already sided with the experts on D3. The case left open, D1, is by contrast the item on which all six judges agree the prediction is incorrect, yet the annotators could not settle whether an unresolved pronominal coalition and a predicate that absorbs the agent name still count as a faithful rendering. Each entry below gives the natural-language input, the gold formula, the model prediction, the verdicts (with any annotator note), and the adjudication outcome. With the two reconciled labels added and the single open case excluded, the human-as-reference comparison rests on $598$ consensus labels.

{\small\sloppy\setlength{\emergencystretch}{3em}
\begin{itemize}[leftmargin=1.3em,itemsep=7pt]
  \item \textbf{D1. Pronominal coalition vs.\ bundled predicate} (\textsc{gpt-5.4}, zero-shot baseline; item \texttt{ex677}).\\
  \emph{Input:} ``It can guarantee that invalid badges will never open the staff entrance, the badge reader.''\\
  \emph{Gold:} \texttt{<<BadgeReader>>G !invalid\_badges\_open\_staff\_entrance}\\
  \emph{Prediction:}\\
  \texttt{<<It>>G(!invalid\_badges\_open\_staff\_entrance\_badge\_reader)}\\
  \emph{Annotator~1} (incorrect): ``prediction uses \texttt{<<It>>} instead of \texttt{<<BadgeReader>>} and bundles \texttt{badge\_reader} into the predicate.''\\
  \emph{Annotator~2}: correct (no note).\quad\emph{All six judges}: incorrect.\\
  The dispute is whether an unresolved pronoun coalition and a proposition that swallows the agent name still count as a faithful rendering. \emph{Adjudication}: \textbf{unresolved}---the annotators could not converge after discussion, so this item is kept as a genuine disagreement and excluded from the human-as-reference set.
  \item \textbf{D2. Reversed \emph{until} operands} (\textsc{qwen-coder-7b}, fine-tuned, zero-shot; item \texttt{ex992}).\\
  \emph{Input:} ``The help desk can guarantee that \ldots the request will remain pending until the missing detail is supplied, the account will remain limited until identity is checked, and the case will remain open until the resolution is confirmed.''\\
  The prediction shares the gold's prefix \texttt{<<HelpDesk>>(X user\_notified \&\& G ticket\_traceable \&\& \ldots)} and differs only in the three \texttt{U} (until) clauses:\\
  \emph{Gold:} \texttt{(request\_pending U missing\_detail\_supplied) \&\& (account\_limited U identity\_checked) \&\& (case\_open U resolution\_confirmed)}\\
  \emph{Prediction:} \texttt{(missing\_detail\_supplied U request\_resolved) \&\& (identity\_checked U account\_limited) \&\& (resolution\_confirmed U case\_open)}\\
  \emph{Annotator~1} (incorrect): ``scoping error: \texttt{U} operands are wrong.''\quad\emph{Annotator~2}: correct (no note).\quad\emph{All six judges}: incorrect.\\
  Every \texttt{U} has its two operands transposed (and one predicate altered), which annotator~1 read as inverting the temporal commitment; the dispute is how strictly operand order must be enforced. \emph{Adjudication}: after discussion the annotators \textbf{agreed the prediction is faithful} (\emph{correct}), overturning the stricter initial label and the six judges' unanimous incorrect verdict.
  \item \textbf{D3. Predicate abbreviation and line-separated conjuncts} (\textsc{mistral-7b}, few-shot baseline; item \texttt{ex462}).\\
  \emph{Input:} ``The pharmacy interface can guarantee that the dispense request remains deferred until the interaction screen is cleared, and the medication safety service can too.''\\
  \emph{Gold:} \texttt{<<PharmacyInterface>>(dispense\_request\_deferred U interaction\_screen\_cleared)}\\
  \hphantom{\emph{Gold:} }\texttt{\&\& <<MedicationSafetyService>>(dispense\_request\_deferred U interaction\_screen\_cleared)}\\
  \emph{Prediction:} the same two coalitions with \texttt{dispense\_request\_deferred} abbreviated to \texttt{request\_deferred} and emitted on two separate lines rather than joined by \texttt{\&\&}.\\
  \emph{Annotator~1}: incorrect (no note).\quad\emph{Annotator~2}: correct (no note).\quad\emph{Judges}: incorrect, except \textsc{gpt-4.1} and the human-aligned \textsc{Llama-3.3-70B} (correct).\\
  The dispute is whether an abbreviated proposition and an implicit (line-break) conjunction preserve the intended meaning. \emph{Adjudication}: after discussion the annotators \textbf{agreed the prediction is faithful} (\emph{correct}); the abbreviation and line-break conjunction were accepted as meaning-preserving, a call that, among the judges, only \textsc{gpt-4.1} and the human-aligned \textsc{Llama-3.3-70B} had made.
\end{itemize}
}

\noindent These cases show that the ``correctness'' of a strategic--temporal translation is not always single-valued at the margins: predicate aliasing, implicit coreference, operand order, and the syntactic marking of conjunction all sit on a boundary where expert judgment legitimately differs. Deliberation resolved two of the three, but in both the experts ultimately accepted a prediction that the LLM judges had rejected, and one case remains genuinely open. This is precisely what motivates our two-tier protocol (reporting the deterministic exact-match floor separately, using six independent judges, and validating them against a human audit) rather than treating any single verdict as ground truth.

\end{document}